\documentclass[a4paper, amsfonts, amssymb, amsmath, reprint, showkeys, nofootinbib, twoside, floatfix, pre,superscriptaddress]{revtex4-2}

\usepackage[utf8]{inputenc}
\usepackage[left=23mm,right=13mm,top=35mm,columnsep=15pt]{geometry} 
\usepackage{amsmath}
\usepackage{amssymb}
\usepackage{esint}
\usepackage[unicode=true, pdfusetitle, bookmarks=true, bookmarksnumbered=false, bookmarksopen=false, breaklinks=false, pdfborder={0 0 1}, backref=false, colorlinks=false]{hyperref}
\usepackage{graphicx}
\graphicspath{ {./images/} }
\usepackage{xcolor}

\makeatletter
\usepackage{makeidx}\usepackage{amsfonts}

\makeatother

\begin{document}
\title{Statistical Mechanics of Inference in Epidemic Spreading}

\author{Alfredo Braunstein}
\affiliation{DISAT, Politecnico di Torino, Corso Duca Degli Abruzzi 24, 10129 Torino}
\affiliation{Collegio Carlo Alberto, P.za Arbarello 8, 10122, Torino, Italy}
\affiliation{Italian Institute for Genomic Medicine, IRCCS Candiolo, SP-142, I-10060 Candiolo (TO), Italy}

\author{Louise Budzynski}
\affiliation{DISAT, Politecnico di Torino, Corso Duca Degli Abruzzi 24, 10129 Torino}
\affiliation{Italian Institute for Genomic Medicine, IRCCS Candiolo, SP-142, I-10060 Candiolo (TO), Italy}
\affiliation{Dipartimento di Fisica, Universit\`a ‘La Sapienza’, P.le A. Moro 5, 00185, Rome, Italy}

\author{Matteo Mariani}
\email{matteo.mariani@polito.it}
\affiliation{DISAT, Politecnico di Torino, Corso Duca Degli Abruzzi 24, 10129 Torino}
\begin{abstract}
    We investigate the information-theoretical limits of inference tasks in epidemic spreading on graphs in the thermodynamic limit. The typical inference tasks consist in computing observables of the posterior distribution of the epidemic model given observations taken from a ground truth (sometimes called planted) random trajectory. We can identify two main sources of quenched disorder: the graph ensemble and the planted trajectory. The epidemic dynamics however induces non-trivial long-range correlations among individuals' states on the latter. This results in non-local correlated quenched disorder which unfortunately is typically hard to handle. To overcome this difficulty, we divide the dynamical process into two sets of variables: a set of stochastic independent variables (representing transmission delays), plus a set of correlated variables (the infection times) that depend deterministically on the first. Treating the former as quenched variables and the latter as dynamic ones, computing disorder average becomes feasible by means of the Replica Symmetric cavity method. We give theoretical predictions on the posterior probability distribution of the trajectory of each individual, conditioned to observations on the state of individuals at given times, focusing on the Susceptible Infectious (SI) model. In the Bayes-optimal condition, i.e. when true dynamic parameters are known, the inference task is expected to fall in the Replica Symmetric regime. We indeed provide predictions for the information theoretic limits of various inference tasks, in form of phase diagrams. We also identify a region, in the Bayes-Optimal setting, with strong hints of Replica Symmetry Breaking. When true parameters are unknown, we show how a maximum-likelihood procedure is able to recover them with mostly unaffected performance.
\end{abstract}

\maketitle

\section{Introduction}
Reconstructing information on epidemic spreading is crucial to develop advanced digital contact tracing strategies in order to mitigate the spreading of an epidemic.
Based on partial information on the states of individuals at given times, the problem consists in reconstructing the posterior distribution on unobserved events, such as the initial state of the epidemic (the source), or undetected infected individuals.
These inverse problems are known to be challenging, even for simple dynamics such as the Susceptible Infectious (SI) model.
Several methods have been proposed to tackle inference problems in epidemics, including Monte Carlo~\cite{antulov-fantulin_statistical_2014, bestvina_infectionmodel_2023, herbrich_crisp_2022}, heuristic \cite{gupta_proactive_2023}, Belief Propagation~\cite{AlBraDaLaZec14, ContactTracing21, braunstein_inference_2016, ghio_bayes-optimal_2023} mean field~\cite{ContactTracing21}, variational~\cite{braunstein2023inference, biazzo_bayesian_2022}, and other~\cite{shah_rumors_2011, pinto_locating_2012} approaches.  Although many of these methods have shown through extensive simulations to reconstruct efficiently some information on the posterior probability distribution in specific graphs sizes and ensembles, a study of the feasibility of inference in epidemic models is still generally lacking. 
A notable exception is given by preprint~\cite{ghio_bayes-optimal_2023} (which appeared while we were finishing the present work). Its main aim is to provide a quantitative study of the feasibility of inference in epidemic spreading on random graphs, in the form of phase diagrams, by means of extensive simulations on finite-size systems. The work focuses on the Bayes optimal setting, and uncovers interesting hints of failure of optimality, that are attributed to finite-size effects.

In this work, we focus on the large size (thermodynamic) limit and use the Replica Symmetric cavity method.  Outside the Bayes optimal regime, we study the performances achieved when hyper-parameters are inferred. We provide a theoretical analysis of inference tasks aiming at reconstructing individuals' trajectories from the partial knowledge of the state of a fraction of individuals at a given observation time \textit{in the thermodynamic limit}, which we also show to be in good agreement with results on moderately large random graphs.
We provide quantitative predictions on the information contained on the posterior probability given the observations, varying the characteristics of the epidemic, of the contact network, and of the observations.
Our approach relies on a study of the properties of the posterior probability measure, for typical contact graphs and realization of the epidemic spreading, using the Replica Symmetric (RS) cavity method. 
We focus in this paper on the simple SI model \cite{SI_description_1994} , but the strategy is general and can be applied to other irreversible spreading process such as the SIR or SEIR models.

To perform this analysis, we need to compute averages of the inference task over realizations of a \textit{planted} epidemic trajectory (the ground truth), from which observations are taken. These observations have thus to be treated as quenched disordered variables (along with the variables needed to describe the contact graph). However, their distribution is non-locally correlated: the past history of the epidemic spreading induces long-range correlations between the state of individuals at the observation time.
While the cavity method is well-suited for models in which variables used to describe the disorder are independent, applying it on a model with long-ranged correlated disorder is instead non-trivial.
To circumvent this difficulty, we devised the following strategy. We separate the \textit{planted} dynamical process in two sets of variables: (a) the \text{transmission delays}, which are independent, and (b) \textit{infection times} and \textit{observations}, which are a deterministic function of other infection times and of transmission delays through a set of local hard constraints. We treat the first set as quenched disorder, while the second set, together with the variables used to describe the inferred trajectory, are treated as dynamical variables. Although \textit{planted} infection times are not truly dynamical variables, their deterministic dependence on the disorder allows us to consider them as such without modifying the probability distribution. This strategy thus effectively transfers correlations out of the quenched variables and into the dynamical ones, allowing a straightforward application of the Cavity Method.

The paper is organized as follows. In section~\ref{sec:formulation} we set up the problem, and present our strategy to adapt the RS formalism to inference in epidemic spreading.
Results are presented in section~\ref{sec:results}. 
We start our analysis in the Bayes-optimal case (section~\ref{subsec:res_Bayes_optimal}), where RS is expected to hold.
We provide quantitative estimates of the feasibility of inference, including Bayes estimators, and the Area Under the ROC Curve (AUC). These RS predictions are in good agreement with the result of message-passing algorithm on large instances.
We identify a region in the Bayes-optimal setting where Belief-Propagation algorithm fails to converge, both on finite-size instances (as already observed in~\cite{ghio_bayes-optimal_2023}) and in the thermodynamic (large-size) limit. This observation is a strong hint for Replica Symmetry Breaking (RSB), and is also confirmed by a failure of Monte-Carlo algorithm performing the inference task in this regime. This result is surprising, as it is often argued that being on the Nishimori line guarantees the absence of dynamic Replica Symmetry Breaking~\cite{LenkaFlorent_StatPhysInf}. However, and although Nishimori's identities are always satisfied in the Bayes-optimal setting, our observations can be explained by the fact that the overlap between planted and inferred trajectories is not necessarily self-averaging in this problem (that is not gauge invariant).
In section~\ref{subsec:results_outsideBO}, we explore regimes outside Bayes-optimal conditions. 
We identify a regime in which neither Belief Propagation nor the iterative numerical resolution of the RS cavity equations converge. This suggests the presence of an RSB transition.
When the parameters of the model are unknown, one can rely on strategies such as Expectation-Maximisation to infer them. These strategies are equivalent to imposing the Nishimori conditions. We provide a quantitative study of an iterative strategy to infer the parameters of the prior in the thermodynamical limit. We show that,  for a large range of the prior's parameters, it is possible to recover similar accuracy than the one of the Bayes optimal case, even when starting from initial conditions that are far from the prior's parameters. These results are in good agreement with simulations in finite systems.

\section{Ensemble Study for Inference in Epidemic Spreading}
\label{sec:formulation}
\subsection{Epidemic Inference}
\paragraph{SI model on graphs.}
We consider the SI model of spreading, defined over a graph $\mathcal{G}=(V,E)$.
At time $t$ a node $i\in V$ can be in two states represented by a variable $x_i^t\in\{S,I\}$.
At each time step, an infected node can independently infect each of its susceptible neighbors $\partial i$ with probabilities $\lambda_{ij}\in[0,1]$. 
\begin{equation}
    P(\underline x)=\prod_{i=1}^N \left[ p(x_i^0)\prod_{t=0}^{T-1}p(x_i^{t+1}|x_i^t,x_{\partial i}^t)\right]
\end{equation}
where $\underline x = \{x_i^t\}$ for $i=1,\dots,N$, $t=0,\dots,T$, and
\begin{align}
    p(x_i^{t+1}=I|x_i^{t},x_{\partial i}^t)&=1-\delta_{x_i^t,S}\prod_{j\in\partial i}(1-\lambda_{ji}\delta_{x_j^t,I})\label{eq:SIx}
\end{align}
The dynamics \eqref{eq:SIx} is irreversible: a given node can only undergo the transition $S\to I$.
Therefore the trajectory in time of an individual can be parameterized by its infection time $t_i$.
\begin{figure}
    \centering
    \includegraphics[width=1.0\columnwidth]{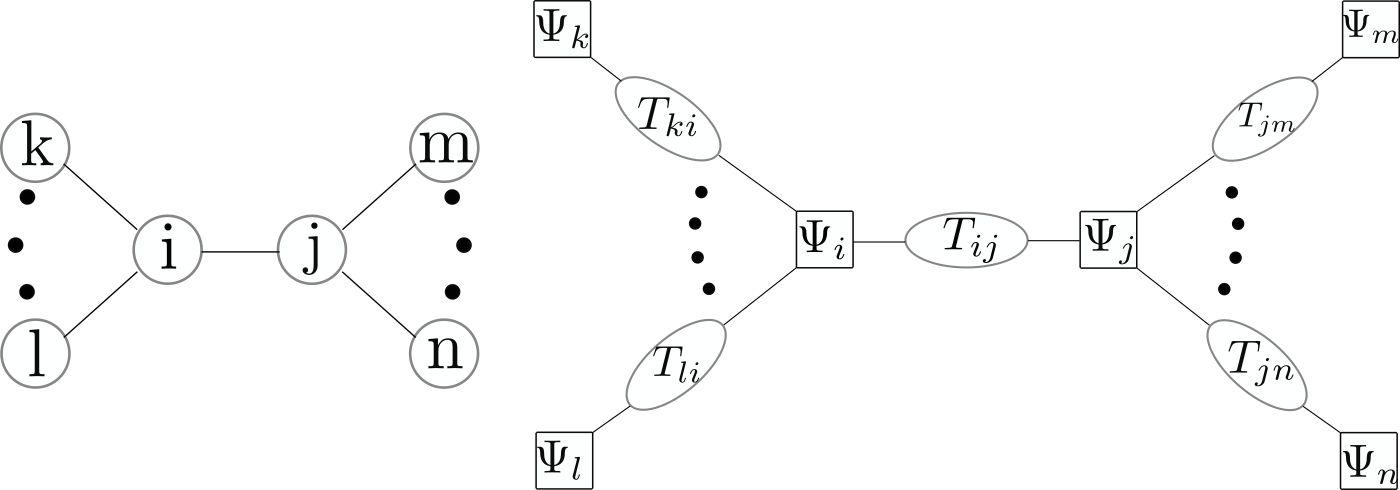}
    \caption{The factor graph construction. On the left there is an example of contact network among individuals. We map this onto the factor graph on the right: for each individual we place a corresponding factor node and for each edge between two individuals we place the super-variable $T_{ij}=(\tau_i^{(j)},\tau_j^{(i)},t_i^{(j)},t_j^{(i)})$ containing the planted and the inferred trajectories of both individuals. This construction increases the complexity of the BP messages, but allows to obtain a disentangled factor graph map (without short loops), which mirrors the contact network.}
    \label{fig:thegraph}
\end{figure}
We assume that a subset of the nodes initiate with an infection time $t_i=0$, i.e. $x_i^0=I$.
A realization of the SI process can be univocally expressed in terms of the independent transmission delays $s_{ij}\in\{1,2, \dots, \infty\}$, following a geometrical distribution $w_{ij}(s)=\lambda_{ij}(1-\lambda_{ij})^{s-1}$.
Once the initial condition $\{x_i^0\}_{i\in V}$ and the set of transmission delays $\{s_{ij},s_{ji}\}_{(ij)\in E}$ is fixed, the infection times can be uniquely determined from the set of equations:
\begin{align}
\label{eq:equation_infected_times}
	t_i=\delta_{x_i^0,S}\min_{j\in\partial i}\{t_j+s_{ji}\}
\end{align}
We assume that each individual has a probability $\gamma$ to be infected at time $t=0$, and we assume for simplicity that the transmission probabilities are site-independent: $\lambda_{ij}=\lambda$ for all $(ij)\in E$.
The distribution of infection times conditioned on the realization of delays and on the initial condition can be written:
\begin{align*}
	P(\underline{t}|\{x_i^0\},\{s_{ij},s_{ji}\})&=\prod_{i\in V}	\psi^*(t_i, \underline{t}_{\partial i}, x_i^0, \{s_{ji}\}_{j\in\partial i}) 
\end{align*}
with $\underline{t}_{\partial i}=\{t_j, j\in\partial i\}$, and where $\psi^*$ enforces the above constraint on the infection times:
\begin{align}
	\label{eq:constraint_infection_times}
	\psi^*=\mathbb{I}[t_i=\delta_{x_i^0,S}\min_{j\in\partial i}\{t_j+s_{ji}\}]
\end{align}
with $\mathbb{I}[A]$ the indicator function of the event $A$.
Once averaged over the transmission delays and over the initial condition, we obtain the following distribution of times:
\begin{align}
	\label{eq:forward_averaged}
	P(\underline{t})&=\prod_{i\in V}\psi(t_i, \underline{t}_{\partial i})
\end{align}
where:
\begin{align*}
	\psi&=\sum_{x_i^0}\gamma(x_i^0)\sum_{\{s_{ji}\}_{j\in\partial i}}\psi^*(t_i, \underline{t}_{\partial i}, x_i^0, \{s_{ji}\}_{j\in\partial i})\prod_{j\in\partial i}w(s_{ji})
 \end{align*}
and $\gamma(x) = \gamma \mathbb{I}[x=I] + (1-\gamma) \mathbb{I}[x=S]$.

\paragraph{Inferring individual's trajectories from partial observations.}
In the inference problem, we assume that some information $\mathcal{O}=\{O_m\}_{m=1,\dots,M}$  is given on the trajectory by the result of $M$ independent medical tests. 
The probability of observations $P(\mathcal{O}|\underline{t})$ factorizes over the set of tests:
\begin{align}
\label{eq:observations}
P(\mathcal{O}|\underline{t})=\prod_{m=1}^M\rho(O_m|t_{i_m}) \ .
\end{align}
Each test gives information about the state $\sigma_m\in\{\text{I,S}\}$ of an individual $i_m$ at a given time $\theta_m\in\{0,1,\dots,T\}$ with a false rate $\textit{fr}\in[0,1]$. We can thus write each observation as $O_m=(i_m,\theta_m,\sigma_m,\textit{fr})$.
The probability of observation conditioned to the infection time is therefore:
\begin{align}
	\label{eq:prob_falserate}
	\rho(O_m|t_{i_m})=(1-\textit{fr})\mathbb{I}[x_{i_m}^{\theta_m}= \sigma_m] + \textit{fr}\mathbb{I}[x_{i_m}^{\theta_m}\neq \sigma_m]
\end{align}
Here for simplicity we assume $\textit{fr}_m\equiv \textit{fr}$ constant which implies identical false positive and negative rates.
Using Bayes rule, the posterior probability of infection times is:
\begin{align}
\label{eq:posterior}
P(\underline{t}|\mathcal{O}) = \frac{P(\underline{t})P(\mathcal{O}|\underline{t})}{P(\mathcal{O})}
\end{align}
with $P(\underline{t})$ given in~(\ref{eq:forward_averaged}).

\paragraph{Bayes optimal setting.}
In the Bayes optimal setting, the parameters $(\lambda,\gamma, \textit{fr})$ of the epidemic spreading process are known in the inference task. This means that the parameters ($\lambda, \gamma, \textit{fr}$) used in the posterior probability~(\ref{eq:posterior}) are the same than the true parameters used to generate the observations.
However in many cases, values of the parameters are unknown, and need to be inferred. In such a case, we denote by ($\lambda^*, \gamma^*, \textit{fr}^*$) (resp. $\lambda^I, \gamma^I, \textit{fr}^I$) the parameters used to generate the observations (resp. to infer the infection times).

\subsection{Ensemble average}
Our objective is to estimate how well observables on the true (or planted) infection trajectory $\underline{\tau}$ are approximated by those of the inferred trajectory $\underline{t}$, which follows the posterior distribution given the observations $\mathcal{O}$. 
We shall characterize the properties of the posterior distribution (\ref{eq:posterior}) on a random ensemble of  contact graphs and realisation of the epidemic spreading.
An instance will be defined by a contact graph $\mathcal{G}$, a ground truth trajectory $\underline\tau$ and a set of observation $\mathcal{O}$ sampled from the distribution $P(\mathcal{O}|\underline{\tau})$.
Three graph ensembles are considered: random regular (RR), Erd\"os-R\'enyi (ER) (defined for example in \cite{mezard_information_2009}), and graph ensemble with a truncated fat tailed (FT) degree distribution. 
We will be interested in the large size limit $n\to\infty$, with $n$ the number of individuals, at fixed degree distribution. In this limit, graph instances of the above-mentioned ensembles are locally tree-like, allowing us to exploit the cavity method in order to determine the typical properties of the measure (\ref{eq:posterior}).

\paragraph{Correlated observations.}
A technical difficulty arises when one tries to apply the cavity method directly to the posterior distribution (\ref{eq:posterior}).
This distribution is defined for a given realisation of the observations $\mathcal{O}=\{O_m\}_{m=1\dots M}$. Observations $\mathcal{O}$ have to be treated as quenched (disorder).
While the cavity method is well-suited for local, independent random disorder, the past history of the epidemic spreading has introduced non-trivial long-range correlations between the observations $\{O_m\}$.
To overcome this difficulty, we rely on the set of hard constraints (\ref{eq:constraint_infection_times}) on the planted times that we recall here: $\tau_i=\delta_{x_i^0,S}\min_{j\in\partial i}\{\tau_j+s_{ji}\}$.
These constraints are expressed in terms of independent random variables: the local delays $s_{ij}$ and the initial-time state $x_i^0$. In fact, knowing these two sets of variables, it is possible to determine each (planted) infection time: for fixed delays $\{s_{ij}, s_{ji}\}_{(ij)\in E}$ and seeds $\{x_i^0\}_{i\in V}$, the planted times are fixed to be the unique solution of the set of constraints (\ref{eq:constraint_infection_times}).
Our strategy is therefore to treat these local variables as disorder (quenched) variables, and consider instead the planted time as constrained dynamical (annealed) variable.
To treat the noise in the observations, we also define the set of error bits $\{\varepsilon_m\}_{m=1,\dots,M}$, with $\varepsilon_m=\mathbb{I}[\sigma_m\neq x_{i_m}^{\theta_m}]$.
We denote by $\mathcal{D}=\{\{x_i^0\}_{i\in V}, \{\varepsilon_m\}_{m=1}^M,\{s_{ij},s_{ji}\}_{(ij)\in E}\}$ the set of all disordered variables. At fixed disorder, the set of planted times $\underline{\tau}$ and observations $\mathcal{O}$ is uniquely determined. 
Averaging over the disordered variables is therefore equivalent to average over the set of observations: with this strategy we can perform the quenched average over correlated observations.
Obviously, the price to pay with this approach is to treat planted times as annealed variables. As a result, the BP messages are over a couple of times $(\tau_i,t_i)$ instead of a single time $t_i$, increasing the complexity in the resolution of the RS equation with population dynamics (see appendix~\ref{app:BP_derivation} for further details).

\paragraph{A graphical model for the joint distribution over planted and inferred trajectories.}
The joint probability of the planted times $\underline\tau$, of the observations $\mathcal{O}=\{O_m\}$ and of the inferred times $\underline{t}$ conditioned on the disorder $\mathcal{D}$ is:
\begin{align}
	P(\underline{t},\mathcal{O},\underline\tau|\mathcal{D}) &= \frac{1}{P(\mathcal{O})}P(\underline\tau|\mathcal{D})P(\mathcal{O}|\mathcal{D},\underline\tau)P(\mathcal{O}|\underline{t})P(\underline{t}|\mathcal{D})
\end{align}
where in the last line we have again noted that the posterior distribution on the inferred times $\underline{t}$ depends only on the observations.
The first term in the product is:
$$
	P(\underline{\tau}|\mathcal{D})=\prod_{i\in V}\psi^*(\tau_i, \underline{\tau}_{\partial i};x_i^0, \{s_{ji}\}_{j\in\partial i})
$$
with $\psi^*$ given in~(\ref{eq:constraint_infection_times}).
The second term in the product is the probability of having observation $\mathcal{O}=\{O_m\}$ given the planted times $\underline{\tau}$ and the disorder $\mathcal{D}$:
$$P(\mathcal{O}|\mathcal{D},\underline{\tau}) = \prod_{m=1}^M\big(
 (1-\varepsilon_m)\delta_{\sigma_m,x_{i_m}^{\theta_m}} + \varepsilon_m(1-\delta_{\sigma_m,x_{i_m}^{\theta_m}})\big).
$$ The third and the fourth terms 
are respectively given in ~(\ref{eq:observations}) and ~(\ref{eq:forward_averaged}). 
Finally, the denominator $$P(\mathcal{O})=\sum_{\underline{t}}P(\underline{t})P(\mathcal{O}|\underline{t})$$ can be seen as a complicated function of the observations $\mathcal{O}$, but since the observations are a deterministic function of the disorder, we will denote it as a function of the latter:
$$
	P(\mathcal{O})=Z(\mathcal{D})
$$

Finally, we obtain the joint probability distribution of planted and inferred times $\underline{\tau}, \underline{t}$ conditioned on the disorder in the form of a graphical model:
\begin{align}
\label{eq:joint_disorder}
\begin{aligned}
	P(\underline{\tau}, \underline{t}|\mathcal{D})=\frac{1}{Z(\mathcal{D})}\prod_{i\in V}&\psi^*(\tau_i, \underline{\tau}_{\partial_i};x_i^0,\{s_{ji}\}_{j\in\partial i}) \\
&\times \psi(t_i,\underline{t}_{\partial i})\xi(\tau_i, t_i;\{\varepsilon_m\}_{i_m=i})
\end{aligned}
\end{align}
with:
\begin{align}
\begin{aligned}
	\xi(\tau_i, t_i;\{\varepsilon_m\}_{i_m=i}) &=\prod_{m:i_m=i} \rho(O_m|t_{i_m}). \\
\end{aligned}
\end{align}
Note that when the error probability is zero: $\textit{fr}=0$ so a single observation reads $O_m=(i_m,\theta_m,\sigma_m,0)$, the error variables are always $\varepsilon_m=0$ (no corruption), and $\rho(O_m|t_{i_m})=\mathbb{I}[\sigma_m=x_{i_m}^{\theta_m}]$. The coupling term $\xi(\tau_i, t_i;\{\varepsilon_m\}_{i_m=i})$ between inferred and planted times in the joint probability becomes:
\begin{align*}
	\xi(\tau_i, t_i)=\prod_{O_m:i_m=i} \mathbb{I}[x_{i_m}^{\theta_m}=\sigma_m] 
\end{align*}

\subsection{Nishimori conditions and Replica Symmetry.}
\label{subsec:Nishimori}
There is a general argument that hints at the absence of replica symmetry breaking in the Bayes optimal case~\cite{iba_nishimori_1999,LenkaFlorent_StatPhysInf}. However, it is not clear that this property holds in general.
The argument is a consequence of the Nishimori conditions, that we reformulate here in our setting.
Consider a given realization of the observations $\mathcal{O}$, and two configurations $\underline{t}_1,\underline{t}_2$ sampled independently from the posterior distribution $P_I(\underline{t}|\mathcal{O})$, where the subscript $I$ refers to the set of hyper-parameters $\lambda^I,\gamma^I,\textit{fr}^I$ used in the inference process.
Let $f(\underline{t}_1,\underline{t}_2)$ be an arbitrary function of two configurations, and $\langle f(\underline{t}_1,\underline{t}_2)\rangle=\sum_{\underline{t}_1,\underline{t}_2}f(\underline{t}_1,\underline{t}_2)P_I(\underline{t}_1|\mathcal{O})P_I(\underline{t}_2|\mathcal{O})$ its average over the posterior distribution. 
Averaging over the observations $\mathcal{O}$, we get:
\begin{align}
\begin{aligned}
	&\mathbb{E}_{\mathcal{O}}[\langle f(\underline{t}_1,\underline{t}_2)\rangle] = \sum_{\mathcal{O}}P_*(\mathcal{O})\langle f(\underline{t}_1,\underline{t}_2)\rangle\\
	&=\sum_{\mathcal{O}}\sum_{\underline{t}_1,\underline{t}_2}P_*(\mathcal{O})f(\underline{t}_1,\underline{t}_2)P_I(\underline{t}_1|\mathcal{O})P_I(\underline{t}_2|\mathcal{O})\\
	&=\sum_{\underline{t}_1,\mathcal{O},\underline{t}_2}P_*(\mathcal{O})\frac{P_I(\underline{t}_1)P_I(\mathcal{O}|\underline{t}_1)}{P_I(\mathcal{O})}P_I(\underline{t}_2|\mathcal{O})f(\underline{t}_1,\underline{t}_2)
\end{aligned}
\end{align}
where in the third line we used the Bayes law for $P_I(\underline{t}_1|\mathcal{O})$, and where the superscript $*$ refers to the set of planted hyper-parameters $\lambda^*,\gamma^*,\textit{fr}^*$ used in the generation of the observables. In the Bayes optimal setting, the two sets of hyper-parameters coincide $(\lambda^*,\gamma^*,\textit{fr}^*)=(\lambda^I,\gamma^I,\textit{fr}^I)$. We therefore obtain the equality:
\begin{align}
\label{eq:Nishimori}
\begin{aligned}
	\mathbb{E}_{\mathcal{O}}[f(\underline{t}_1,\underline{t}_2)] &=\sum_{\underline{\tau},\mathcal{O},\underline{t}}P_*(\underline{\tau})P_*(\mathcal{O}|\underline{\tau})P_*(\underline{t}|\mathcal{O})f(\underline{\tau},\underline{t})\\
	&=\mathbb{E}[f(\underline{\tau},\underline{t})]
\end{aligned}
\end{align}
with $\underline{\tau}$ a planted configuration, and $\underline{t}$ a random sample of the posterior.
The quenched average $\mathbb{E}_{\mathcal{O}}$ is over the set of observations $\mathcal{O}$. Note that in our case, we average instead over the set of disordered variables $\mathcal{D}$: as explained above this is equivalent to average over observations $\mathcal{O}$, since $\mathcal{O}$ is a deterministic function of $\mathcal{D}$.
The equality~(\ref{eq:Nishimori}) is true in particular when the function $f$ is taken to be the overlap between two configurations. 
In words, it states that the average of the overlap $q$ between two configurations sampled independently from the posterior, is equal to the average of the overlap $q^*$ between the planted configuration and a random sample from the posterior. 
Applying this equality to higher moments of the overlap, it is actually possible to show that the two distributions are equal.
In models with gauge invariance, such as the planted spin glass studied in \cite{iba_nishimori_1999, LenkaFlorent_StatPhysInf}, the overlap $q^*$ coincides with the magnetization. It can be argued that the magnetization is a self-averaging quantity, and therefore that the overlap $q$ is also self-averaging on the Nishimori line. This argument allows to conclude that the probability distribution of the overlap $q$ is trivial, and therefore that there is no replica symmetry breaking phase in the Bayes optimal setting. In the case of epidemics,
it is less clear that the overlap $q^*$ between the planted configuration and a random sample of the posterior is a self-averaging quantity. In fact, we
observed a region (for small seed probability $\gamma$ and large transmission rate $\lambda$), which presents signs of a failure of optimality, signalled by a lack of convergence of Belief-Propagation in the thermodynamic limit (see section \ref{sec:results}, where we conjecture that this is due to replica symmetry breaking). A similar observation is made in \cite{ghio_bayes-optimal_2023} on finite size instances.

\subsection{Belief-Propagation equations for the joint-probability}
\label{subsec:BP_for_joint}
The factor graph associated with the probability distribution~(\ref{eq:joint_disorder}) contains short loops which compromise the use of BP. 
In order to remove these short loops, we introduce the auxiliary variables $(\tau_i^{(j)},\tau_j^{(i)},t_i^{(j)},t_j^{(i)})$ on each edge $(ij)\in E$ of the factor graph, which are the copied times $\tau_i^{(j)}=\tau_i$, and $t_i^{(j)}=t_i$ for all $j\in\partial i$.
Let $T_{ij}=(\tau_i^{(j)},\tau_j^{(i)},t_i^{(j)},t_j^{(i)})$ be the tuple gathering the copied times on edge $(ij)\in E$.
The probability distribution on these auxiliary variables is:
\begin{align}
\label{eq:prob_auxiliary}
	P(\{T_{ij}\}_{(ij)\in E}|\mathcal{D}) &= \frac{1}{\mathcal{Z}(\mathcal{D})}\prod_{i\in V}\Psi(\{T_{il}\}_{l\in\partial i};\mathcal{D}_i) 
\end{align}
where $\mathcal{D}_i=\{\{s_{li}\}_{l\in\partial i},x_i^0, \{\varepsilon_m\}_{i_m=i} \}$ is the disorder associated with vertex $i\in V$, and with:
\begin{align}
\begin{aligned}
	\Psi(\{T_{il}\}_{l\in\partial i};&\mathcal{D}_i) =\\
 &\xi(\tau_i^{(j)},t_i^{(j)};c_i)\psi^*(\tau_i^{(j)},\underline{\tau}_{\partial i}^{(i)};\{s_{li}\}_{l\in\partial i},x_i^0)\times\\
 \times&\psi(t_i^{(j)},\underline{t}_{\partial i}^{(i)})\prod_{l\in\partial i\setminus j}\delta_{t_i^{(j)},t_i^{(l)}}\delta_{\tau_i^{(j)},\tau_i^{(l)}}
 \end{aligned}
\end{align}
where $j\in\partial i$ is a given neighbour of $i$. The factor graph associated with this probability distribution now mirrors the original graph $\mathcal{G}=(V,E)$ of contact between individuals, as shown in Figure \ref{fig:thegraph}.
The variable vertices live on the edges $(ij)\in E$, and the factor vertices associated with the function $\Psi$ live on the original vertex set $V$. 
We introduce the Belief Propagation (BP) message $\mu_{i\to \Psi_j}$ on each edge $(ij)\in E$ as the marginal probability law of $T_{ij}$ in the amputated graph in which node $j$ has been removed.
The set of BP messages obey a set of self-consistent equations:	
\begin{align}
\label{eq:BP_equations}
\begin{aligned}
	\mu_{i\to \Psi_j}(T_{ij}) &= \\ 
=\frac{1}{z_{\Psi_i\to j}}&\sum_{\{T_{il}\}_{k\in\partial i \setminus j}}\Psi(\{T_{il}\}_{l\in\partial i};\mathcal{D}_i)\prod_{k\in\partial i \setminus j}\mu_{k\to\Psi_i}(T_{ik})
\end{aligned}
\end{align}	
were $z_{\Psi_i\to j}$ is a normalization factor. 
These equations are exact when the contact graph $\mathcal{G}=(V,E)$ is a tree. In practice, the BP method is also used as a heuristic on random sparse instance.
Introducing a horizon time $T$, the random variable $T_{ij}=(\tau_i^{(j)},\tau_j^{(i)}, t_i^{(j)},t_j^{(i)})$ lives in a space of size $O(T^4)$.
We see in appendix~\ref{app:BP_derivation} how to simplify the BP equations \eqref{eq:BP_equations}, and obtain a set of equivalent equations defined over modified BP messages living in a smaller space.

\subsection{Estimators}
\label{sec:estimators}
To quantify the feasibility of inference tasks, some estimators are defined in this paragraph and studied in the Results section~\ref{sec:results}. In view of that, it is useful to define $P_{i,t}(x_i^{*,t},x_i^t)$ as the marginal probability of having the planted state $x_i^{*,t}$ and the inferred state $x_i^t$ of one individual $i\in V$ at a given time $t\in\{0,1,\dots,T\}$.
\paragraph{Maximum Mean Overlap}
The overlap at a given time $t$ between the planted configuration $\underline{x}^{*,t}$ and an estimator $\underline{\hat{x}}^t$ is $O_t(\underline{x}^{*,t},\underline{\hat{x}}^t) = \frac{1}{N}\sum_{i=1}^N\delta_{x_i^{*,t},\hat{x}_i^t}$. 
In the inference process, on a given instance, the planted configuration is not known, and the best Bayesian estimator is obtained by assuming that $\underline{x}^{*,t}$ is distributed according to the posterior distribution. The best estimator of the overlap $\underline{\hat{x}}^{t,{\rm MMO}}$ is the one maximising the overlap averaged over the posterior:
\begin{align}
    {\rm MO}_t(\underline{\hat{x}}^t) = \sum_{\underline{x}^t}P(\underline{x}^t|\mathcal{O})O_t(\underline{x}^t,\underline{\hat{x}}^t) \ ,
\end{align}
which is achieved for $\hat{x}_i^{t,{\rm MMO}}={\rm argmax}_{x_i^t}(P_{i,t}(x_i^t|\mathcal{O})$.
The overlap $O_t(\underline{x}^{*,t}, \underline{\hat{x}}^{t,{\rm MMO}})$ provides a quantitative estimation of the accuracy of the Maximum Mean Overlap estimator $\underline{\hat{x}}^{t,{\rm MMO}}$.
We compute this quantity, averaged over the graph ensemble, and the realisation of the planted configurations and of the observations: $$\mathbb{E}_{\mathcal{G},\mathcal{D}}[O_t(\underline{\hat{x}}^{*,t}, \underline{\hat{x}}^{t,{\rm MMO}})] \ .$$ 
Note that in our formalism, we have access the marginal probability over planted and inferred configurations, conditioned on the disorder $\mathcal{D}$: $P_{i,t}(x_i^{*,t}, x_i^t|\mathcal{D})$. 
However, as previously explained, fixing the disorder variables is sufficient to fix the planted configuration and the observations $\mathcal{O}$.  
\paragraph{Minimum Mean Squared Error}
We also consider the squared error (SE) at a given time $t$ between the planted configuration $\underline{x}^{t,*}$ and an estimator $\underline{\hat{x}}^t$: $${\rm SE}_t(\underline{x}^{*,t},\underline{\hat{x}}) = \frac{1}{N}\sum_{i=1}^N(x_i^{*,t}-x_i)^2.$$ 

As for the overlap, the best Bayesian estimator for the squared error $\underline{\hat{x}}^{t,{\rm MMSE}}$ is the one minimizing the squared error averaged over the posterior:
\begin{align}
    {\rm MSE}_t(\underline{\hat{x}}^t)=\sum_{\underline{x}^{t}}P(\underline{x}^t|\mathcal{O})SE(\underline{x}^t,\underline{\hat{x}}^t)
\end{align}
which is achieved for $\hat{x}_i^{t,{\rm MMSE}}=\sum_{x_i^t}P_{i,t}(x_i^t|\mathcal{O})x_i^t$.
We compute the average, over the graph ensemble and the disorder, of the squared error between the planted configuration and the MMSE estimator:
$$
    \mathbb{E}_{\mathcal{G},\mathcal{D}}[{\rm SE}_t(\underline{\hat{x}}^{*,t}, \underline{\hat{x}}^{t,{\rm MMSE}})] \ .
$$
\paragraph{Area Under the Curve (AUC)}
On a single instance, the receiver operating characteristic (ROC) curve is computed as follows. At a fixed time $t$, one computes for each individual its marginal probability $P_i(x_i^t=I|\mathcal{O})$. For a given threshold $\rho\in[0,1]$, the true positive rate TPR($\rho$) (resp. false positive rate FPR($\rho$)) is the fraction of positive (resp. negative) individuals with $P_i(x_i^t=I|\mathcal{O})\geq \rho$. The ROC curve is the parametric plot of TPR($\rho$) versus FPR($\rho$), with $\rho$ the varying parameter. Note that the FNR (and therefore the ROC curve) is undefined when all individuals are infected (all positive).
The area under the curve (AUC) can be interpreted as the probability that, picking one positive individual $i$ and one negative individual $j$, their marginal probabilities allows to tell which is positive and which one is negative, i.e. that $P_i(x_i^t=I|\mathcal{O}) > P_j(x_j^t=I|\mathcal{O})$.
This allows us to compute the AUC under the Replica Symmetric formalism.

\subsection{Inferring the hyper-parameters}
The prior parameters, with which the planted epidemic is generated, might not be accessible in the inference process. In those cases, we propose to infer them from the observations by approximately maximizing $P(\mathcal{O}|\lambda^I,\gamma^I)$ at fixed observations $\mathcal{O}$. In this section we provide an upper bound to the feasibility of parameters inference. It is an upper bound because the process is described for the ensemble case, i.e. for an infinite sized contact graph. This means that also the number of observations is infinite. The idea is to find the most typical parameters $\gamma^I,\lambda^I$ for the given set of observations. For inferring $\gamma$ we use the Expectation Maximization (EM) method, an iterative scheme which consists in separating the optimization process in two steps:
\begin{enumerate}
    \item At fixed BP messages, the update for $\gamma$ at $k_{th}$ iteration reads: 
    \begin{equation}
    \label{eq:EM_general}
        \gamma_{k}=\arg\max_{\gamma}\left\langle \log P\left(\underline{t},O|\gamma\right)\right\rangle _{\{\mu\}_{k}},
    \end{equation} where ${\{\mu\}_{k}}$ is a shorthand for the set of all BP messages at $k_{th}$ iteration.
    \item At fixed $\gamma_k$, the messages are updated with BP equations. 
\end{enumerate}
To understand \eqref{eq:EM_general} we recall the definition of the variational free energy: 
\begin{equation}
    F[Q](\gamma) := -\left\langle \log P(\underline{t},\mathcal{O}|\gamma)\right\rangle _{Q}+\left\langle \log Q(\underline{t})\right\rangle _{Q} 
\end{equation}
The posterior distribution $P(\underline{t}|\mathcal{O},\gamma)$ can be shown to be the distribution $Q$ which minimizes $F$ (see for example \cite{parisi_statistical_1998}). If we evaluate averages with fixed BP messages, then the dependency of $F$ on $\gamma$ is only on the first addend of the right hand side. Then the optimization on $\gamma$ reduces to equation \eqref{eq:EM_general}. 
By setting to zero the first derivative of \eqref{eq:EM_general} w.r.t. $\gamma$ we have, for the $k_{th}$ iteration:
\begin{equation}
\label{eq:EM_gamma}
    \gamma_k=\frac{1}{N}\sum_{i\in V}p^{I,k}_{i}(t_i=0|\mathcal{O})
\end{equation} 
Where $p^{I,k}_{i}(t_i=0|\mathcal{O})$ is the posterior probability at $k_{th}$ iteration of EM for individual $i$ to have infection time equal to 0 (i.e. to be the patient zero). The procedure we propose to include the inference of patient zero probability is therefore to simply update $\gamma^I$ with  equation \eqref{eq:EM_gamma} at every sweep of BP update on the population.
We could write equations for EM in $\lambda$, but they would be more involved. We opted therefore to simply perform a gradient descent (GD) on the Bethe Free energy.
For the epidemic propagation on graph, the Bethe free energy can be related to the normalization of the BP messages  and BP beliefs, respectively addressed as $z_{\Psi_{i}\to j}$ and $z_{\Psi_{i}}$:
$$F_{\text{Bethe}}=\sum_{i\in V}\left[\left(\frac{1}{2}|\partial i|-1\right)\log z_{\Psi_{i}}-\frac{1}{2}\sum_{j\in\partial i}\log z_{\Psi_{i}\to j}\right]$$
Overall, the inference of prior hyper-parameters is done by alternating one sweep of BP update of the messages at fixed parameters $(\gamma^I,\lambda^I)$ with one step of EM for $\gamma$ and one step of GD for $\lambda$.

\section{Results}
\label{sec:results}

In this section we explore the performance of inference tasks for the SI model. We consider two different regimes:
\begin{enumerate}
    \item When the parameters of the prior distribution are known in the inference process. (Bayes optimal setting)
    \item When the parameters of the prior distribution are not known, and are inferred.
\end{enumerate}
The results are obtained by randomly initializing a population of BP messages and by iterating a population dynamics algorithm, as described in appendix~\ref{sec:RS_equations}. The algorithm stops when the BP messages satisfy a simple convergence criterion on the marginals, which must not fluctuate more than the square root of size of the population. If convergence criterion is not reached, the algorithm stops after a fixed number of \textit{maxiter} sweeps (typically we set the population size $N\sim 10^4$ and the total number of sweeps \textit{maxiter} $=100$). Convergence is (almost always) reached when the prior is known (or inferred), except for a rather interesting and unexpected regime which is discussed later on. The algorithm shows non-converge zones, as expected, also when the prior is not known.

\subsection{Results in the Bayes-Optimal case.}
\label{subsec:res_Bayes_optimal}
\begin{figure}
    \centering
    \includegraphics[width=0.45\textwidth]{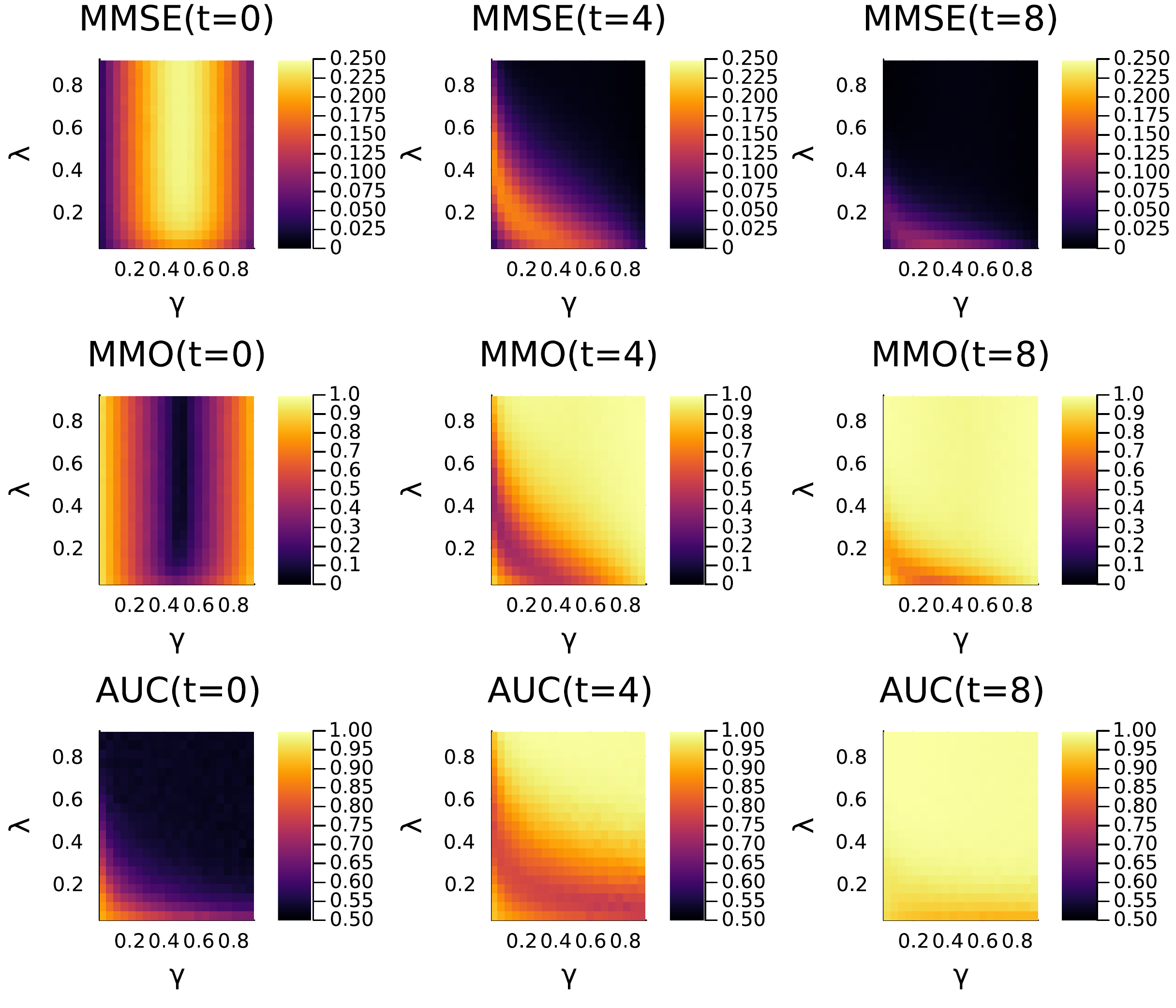}
    \caption{Several measures: (first row: MMSE, second row: MMO, third row: AUC) quantifying the hardness of epidemic inference, as a function of patient zero probability $\gamma$, and infection probability $\lambda$. Each column corresponds to three different times at which the quantities are computed (from left to right: initial time $t=0$, intermediate time $t=4$, and final time $t=T=8$). The three measures display the same behaviour, except for the initial time, when AUC is able to capture for high values of $\lambda$ and $\gamma$ that observations are not informative enough. Notice that MMSE quantifies the error in inferring individual's states, so it has a flipped behaviour with respect to the other quantities (MMO and AUC are high when inference performance is good). These results were obtained for ER graph ensemble with average degree $3$.}
    \label{fig:comp_measures}
\end{figure}
In this paragraph, we study several measures and estimators that quantify the hardness of inference, varying epidemic parameters (transmission probability $\lambda$, patient zero probability $\gamma$), but also the fraction of observed individuals. We compare the Minimum Mean Squared Error (MMSE), the Maximum Mean Overlap (MMO), and the Area Under the ROC (AUC) (see section \ref{sec:estimators} for their definition), and the Bethe Free Energy (Fe) associated with the posterior distribution $P(\underline{t}|\mathcal{O})$. 
In Figure \ref{fig:comp_measures}, we fix the fraction of unobserved individuals (dilution) to $\textit{dil}=0.5$ (half the individuals are observed). We set the observation time at final time $T=8$, and explore the space $(\gamma,\lambda)$.
MMSE, MMO and AUC are computed at three different times (initial time $t=0$, intermediate time $t=4$, and final time $t=T=8$).
We can see that MMSE and MMO show the same behaviour at all times. 
For very low infection probability $\lambda$, and patient zero probability $\gamma$, we see that MMSE is low (while MMO and AUC are high), meaning that the information contained in the inferred posterior distribution allows to recover the planted configuration with good accuracy. In this regime, typically seeds are surrounded by a small neighborhood of infected individuals, well-separated from the other seeds, making inference task easy. For high values of patient zero probability $\gamma$ and infection $\lambda$, instead, the population becomes completely infectious in few time steps. Also in this regime, all the estimators show great performance for intermediate (t=4) and final times, because the posterior marginals assign to every individual a probability $1$ of being infectious. 
The hard regime is for intermediate values of $\gamma, \lambda$.

Note also that at $t=0$, MMSE (respectively MMO) is low (resp. high) for high values of $\gamma$. 
However, this does not mean that inference performance is good in this regime. Indeed for large $\gamma$, the majority of individuals are patients zero, and the other individuals are likely to be infected before the observation time $T$. Therefore, the observations are (almost) all positive, making impossible to distinguish the patients zero from the ones infected at later time. Thus, MMSE at time $t=0$ is low because the marginal posteriors give high probability of being infected at $t=0$, independently of the transmission rate $\lambda$. However, the (few) non-patients zero will remain undetected.
A quantity that is sensible to this problem is the AUC, which at time $t=0$ has in fact a different behaviour with respect to the other measures. 
Another (slightly) different quantity, for example, is the AUC evaluated only on non observed individuals. 
\begin{figure}[b]
    \centering
    \includegraphics[width=0.45\textwidth]{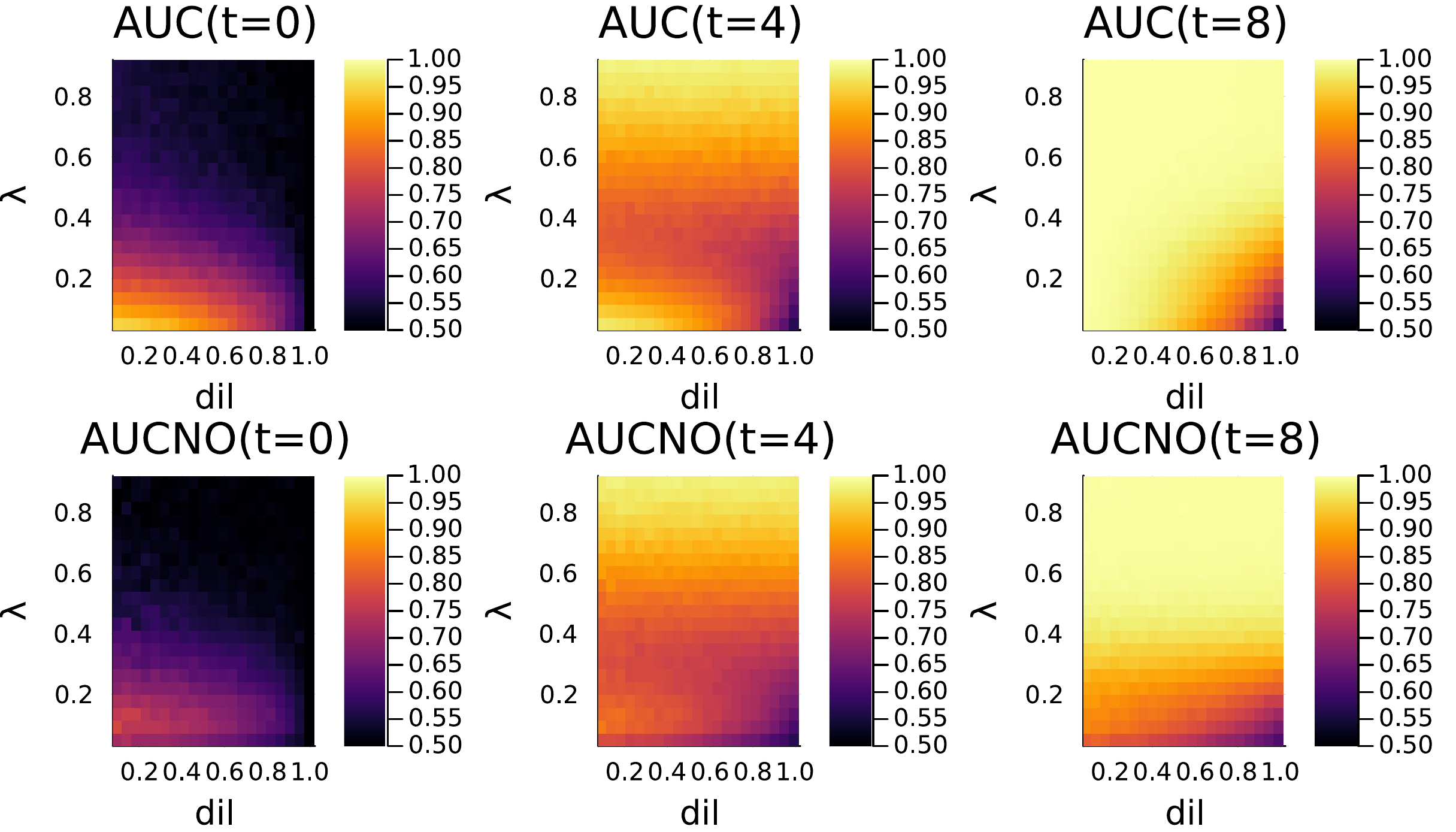}
    \caption{A comparison between AUC evaluated on all individuals (AUC, first row) VS only unobserved individuals (AUCNO, second row). The two measures have a very similar but not identical behaviour. In particular, at low dilution (many observations) the AUCNO is systematically smaller than the AUC as expected. This difference is more pronounced for low values of $\lambda$, and at intermediate and final time. These results are for ER graph ensemble, with average degree $3$. We remark here that AUC is not $0.5$ for dilution equal to $1$. In fact, ER graphs are heterogeneous (with a Poisson law degree distribution). This implies that some information about the probability of infection of each node is contained in the graph itself. For example, the most connected nodes have highest probability of being infected. This allows to achieve some reconstruction also without any observation (dil=$1$)}
    \label{fig:AUCvsAUCNO}
\end{figure}
When many observations are done, the AUC is dominated by the observed individuals. Thus, evaluating AUC only on non observed individuals (AUCNO) can be a useful tool to understand the predicting power of the algorithm on individuals for which no information is given. To see the difference between AUC and AUCNO, we fix the patient zero probability $\gamma=0.1$, and study these two measures as a function of the infection probability $\lambda$ and the observations dilution \textit{dil} (i.e. the fraction of unobserved individuals), see Figure \ref{fig:AUCvsAUCNO}. We see that the two estimators behave differently, for example at the intermediate time $t=4$, for low dilution (i.e. many observations) and low transmission rate $\lambda$. In this regime, there are only few infected individuals at the observation time (low $\gamma$ and $\lambda$). While AUC is close to $1$, AUCNO is low, indicating that it is actually hard to find who are the \textit{unobserved} infected individuals. 
\begin{figure}
    \centering
    \includegraphics[width=0.4\textwidth]{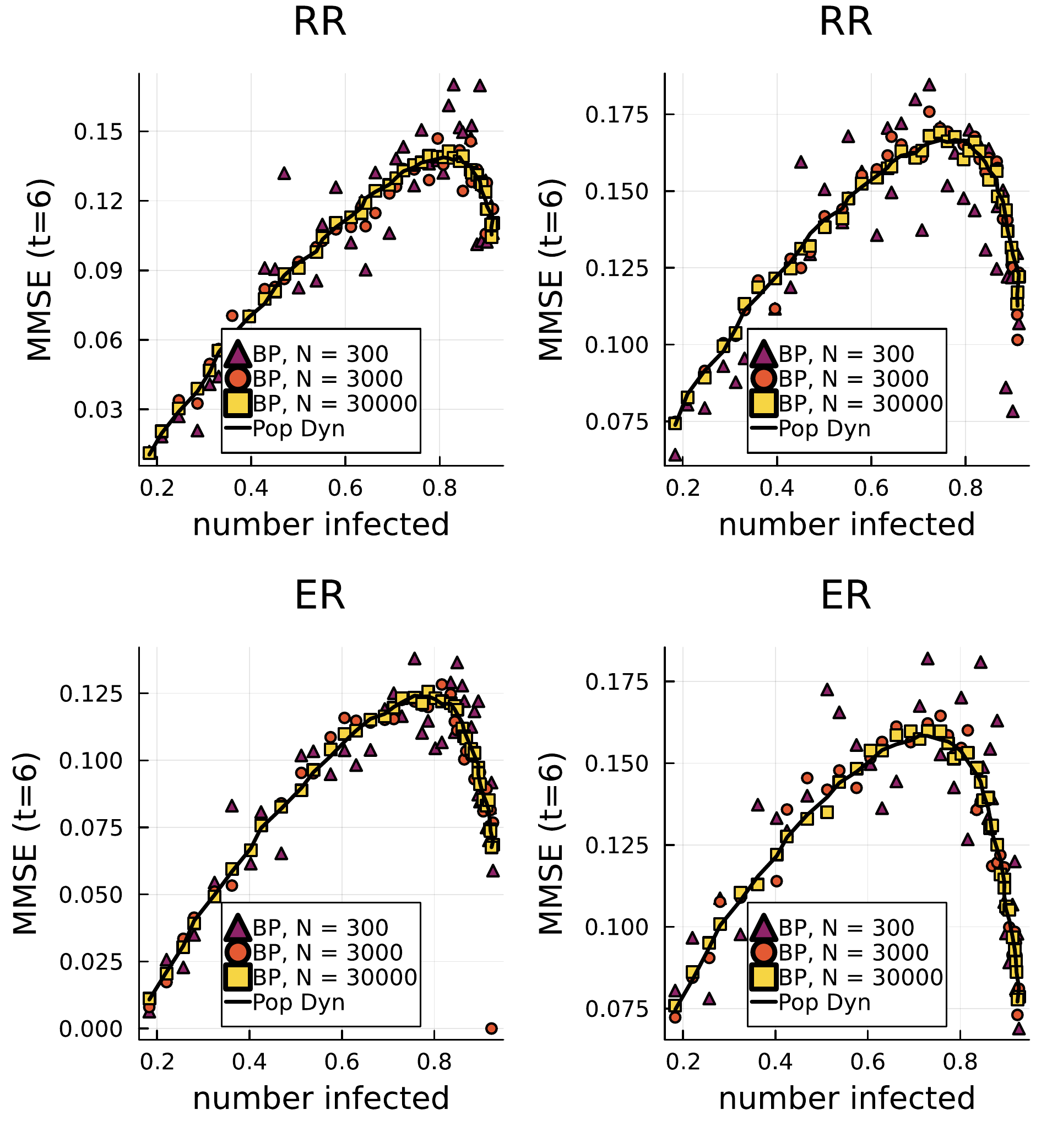}
    \caption{The comparison between Sibyl algorithm (BP) for a single instance of $N=300$ (triangles), $N=3000$ (dots), and $N=30000$ (squares) individuals, and the ensemble results obtained in the thermodynamic limit with population dynamics (black solid line). The plots show the MMSE at intermediate time $t=6$, as function of the number of infected at final time $T=8$, which is a function of infection parameters $\gamma$ and $\lambda$. For this plot the patient zero probability is fixed at $\gamma=0.15$. The first row represents the MMSE for Random Regular graphs (degree $3$) while the second row is for Erd\"os-R\'enyi (ER) with average degree $3$. Each column instead is associated with a value of observations dilution \textit{dil}: the first column is for $\textit{dil}=0$ (all observed) while the second is for $\textit{dil}=0.5$. We see a very good agreement, that increases with the size of the the single instance contact graph, and that we checked to persist in the other observables (MMO and AUC).}
    \label{fig:sib_vs_ens}
\end{figure}

\paragraph{Comparison with finite-size instances.}
It is natural to wonder whether our ensemble results obtained in the thermodynamic limit, with the RS cavity method, are consistent with large finite-size single instances. To check this point, we initialized large sized ($N=30000$) graphs, and simulated discrete-time epidemic spreading and observation protocol. We used Sibyl \cite{AlBraDaLaZec14}, which is a Belief Propagation algorithm for calculating the posterior marginals in single instance problems. We computed the MMSE, and we compared it with the RS predictions. An example is shown in Figure \ref{fig:sib_vs_ens}, showing a good agreement between the RS predictions and the results on a single large instance.
\begin{figure}
    \centering
    \includegraphics[width=0.45\textwidth]{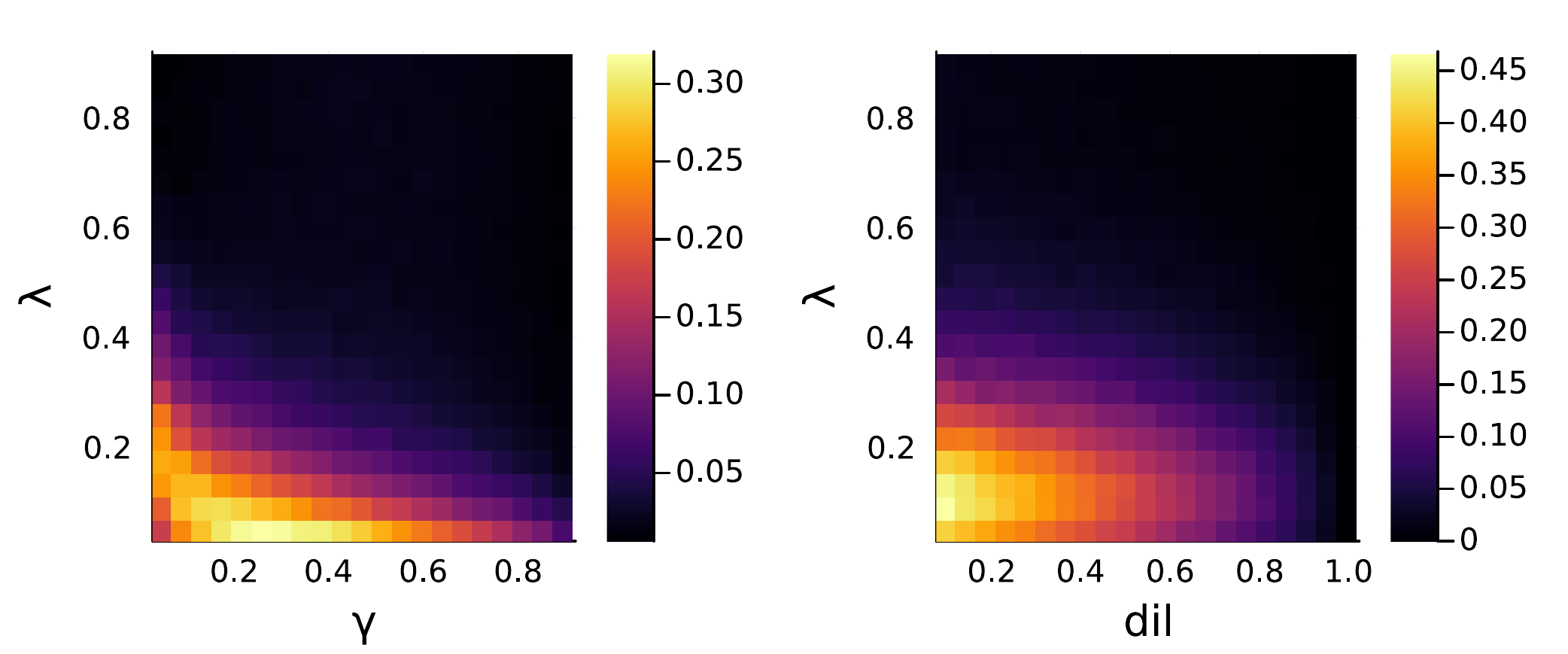}
    \caption{Free energy profile for two different regimes. Left panel: as a function of patient zero probability $\gamma$ and infection probability $\lambda$, at fixed dilution $\textit{dil}=0.5$. Right panel: as a function of observations dilution $\textit{dil}$ and $\lambda$, at fixed patient zero probability $\gamma=0.1$. The black part of the plot corresponds to the regimes in which observations do not bring any information, i.e. $F\simeq 0$. This happens obviously at $\textit{dil}=1$ because no observation is done. However, also for non-zero dilution, the free energy can be zero. When the infection transmission $\lambda$ is high enough, in fact, all the individuals are infected at final time with probability almost equal to 1. Since for this plot the observations are performed at final time, then they all simply register the infectiousness of each individual, factually carrying no information with them. Only in the intermediate regimes, i.e. when the numbers of infectious and susceptible individuals are comparable to each others, observations carry information. In this regime the free energy is non-zero and inference is non trivial. The graph ensemble analyzed here is Erd\"os-R\'enyi  with average degree $3$.}
    \label{fig:free_energies}
\end{figure}

\paragraph{Information contained in the observations.}
All the observables described so far are time-dependent quantities, giving an estimation on how easy/hard it is to infer the planted individual states $x_i^{*,t}$ at a given fixed time $t$. It is useful to define a time-independent observable, which gives a general overview of the inference process. We opted to study the Bethe Free Energy $F=-\log Z(\mathcal{D})=-\log P(\mathcal{O})$, which can be expressed in terms of BP marginals (see its expression in appendix~\ref{subsec:BetheFreeEnergy}). 
It is the free energy associated with the posterior distribution:
\begin{equation}
    P(\underline{t}|\mathcal{O})=\frac{P(\underline{t},\mathcal{O})}{P(\mathcal{O})},    
\end{equation}
where 
$$P(\mathcal{O})=\sum_{\underline{t}} P({\underline{t}},\mathcal{O})=\sum_{\underline{t}} P(\mathcal{O}|{\underline{t}})P({\underline{t}}).$$
$F=-\log P(\mathcal{O})$ quantifies how informative the observations $\mathcal{O}$ are: the quantity $P(\mathcal{O})$ is the sum of trajectories (weighted with their prior probability) which are compatible with the observation constraints $\mathcal{O}$. Observations reduce the space of the possible trajectories; as an extreme example, if we observed (noiselessly) every individual at every time, the space of possible trajectories would collapse on a single trajectory (the planted solution). The plots in Figure \ref{fig:free_energies} show the free energy in the two different regimes discussed above. The free energy is obviously $0$ for $\textit{dil}=1$ (no observation). 
However, it is close to $0$ in other cases too, e.g. for high values of $\lambda$. For those values, the infection spreads very fast. As a consequence, at final time all the individuals are infected. Thus, since the observations are taken at final time, they do not bring valuable information on the planted trajectory: they will always register a positive (infected) result for all individuals at final time.
In other words, all trajectories sampled from the prior are compatible with the observation that all individuals are infected at time $T$.
Note however that inference can be easy in this regime, as it can be checked comparing Figure~\ref{fig:free_energies} with Figures~\ref{fig:comp_measures} and \ref{fig:AUCvsAUCNO} (for times $t=4$ and $t=8$). In this regime, although observations are not informative, the prior is concentrated on few trajectories (that are compatible with all individuals being infected at times $t=4$ and $t=8$), making inference task trivial.
The interesting (and hard) regimes are at intermediate/low values of $\gamma$ and $ \lambda$ and for non-zero dilution. In this regime, although observations are informative ($F$ positive) the prior is not concentrated on few trajectories, making the inference a non-trivial task.

\paragraph{More graph ensembles.}
The analysis shown so far has been performed on Erd\"os-R\'enyi graphs. 
To study if and how inference performance is affected by the graph structure, we compared the results on three families of graphs:
\begin{enumerate}
    \item The Random Regular (RR) ensemble, where each node has the same degree $d$;
    \item The Erd\"os-R\'enyi (ER) ensemble. In the large size limit, the degree distribution is a Poisson distribution of average $d$;
    \item A (truncated) \textit{fat tailed} (FT) ensemble of graphs, with a degree distribution $p(d)=\frac{1}{Z}\frac{1}{d^2+a}$ for $d\in[d_{min},d_{max}]$ and $p(d)=0$ if $d\notin[d_{min},d_{max}]$. The quantity $Z$ is the normalization of the distribution and the parameter $a$ can be fixed by fixing the average degree.
\end{enumerate}
The choice for the third graph ensemble allows for the existence of highly connected nodes, while still being handled by Belief Propagation (BP), since the distribution of the degree is truncated to a finite maximum value $d_{max}$.
\begin{figure}
    \centering
    \includegraphics[width=0.45\textwidth]{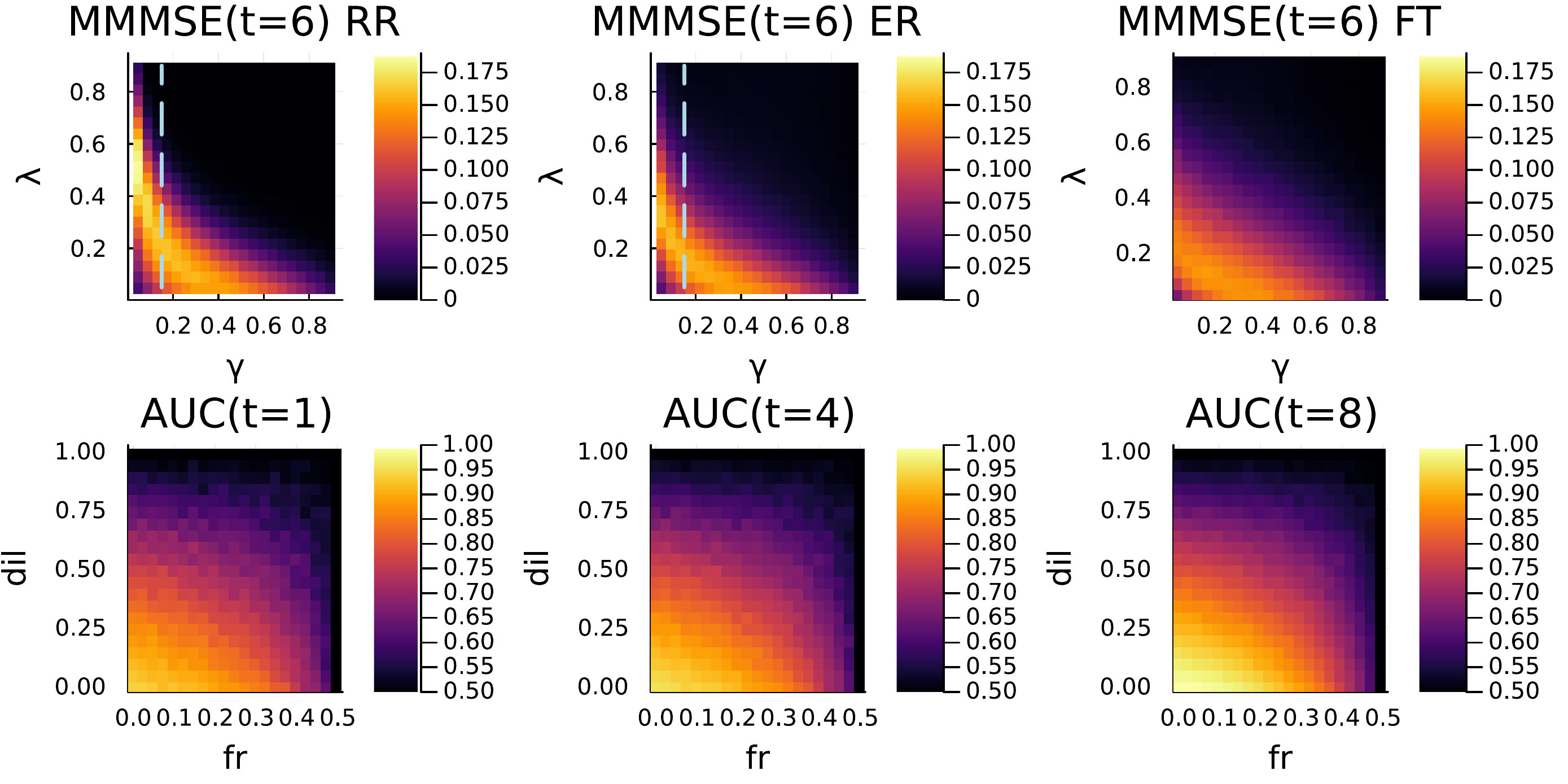}
    \caption{Comparing feasibility of inference for different graph ensembles and for nonzero observation noise. \textit{First row:} the plots show the MMSE at time $t=6$, with observations made at final time $T=8$, as functions of the patient zero probability $\gamma$ and the infection probability $\lambda$. The three plots are (from left to right) for Random Regular (RR), Erd\"os-R\'enyi (ER) and Fat Tailed (FT) graph ensembles. The average degree is fixed to $3$ for the three ensembles examined. It can be seen that the profiles share similarities, but the more the degree distribution widens (from RR to ER to FT), the flatter is the MMSE. This is due to the presence (or absence) of high-degree nodes. In RR ensemble, all nodes have the same degree, so for example we see that inference is more difficult at low values of $\gamma$ and values of $ \lambda$. In this region, the presence of highly connected nodes simplifies inference because they (and their neighbours) will probably be infectious at time t=6.
    The dashed lines correspond to the cases studied in Figure \ref{fig:sib_vs_ens} at dilution $0.5$. The only difference is in the $y$-axis, which is $\lambda$ for this plot and the number of infected for the plots in Figure \ref{fig:sib_vs_ens}. 
    \textit{Second row:} the AUC as function of observations' dilution (\textit{dil}) and false rate (\textit{fr}). The AUC  decreases with \textit{fr} and \textit{dil}. The false positive rate and false negative rate are assumed to be the same. The patient zero probability is fixed to $\gamma=0.03$ and the infection probability is $\lambda=0.03$. The ensemble graph is Random Regular with degree $11$.}
    \label{fig:graphs_fr}
\end{figure}
In Figure \ref{fig:graphs_fr} (first row), we compare the Minimum Mean Squared Error (MMSE) at time $t=6$ for the three ensembles of graph. The average degree is fixed to $3$ in all three graph ensembles. 

\paragraph{Noise in observations.}
When noise affects individuals' observations, the inference results get typically worse. This can be seen in Figure \ref{fig:graphs_fr} (second row), where we studied the AUC as function of observations dilution and noise (false rate, \textit{fr}). For false rate equal to $0.5$, the observations carry no information, since they are wrong half of the time. This is identical to set dilution to $\textit{dil}=1$, i.e. not performing any observation. For intermediate values, we see that increasing false rate and/or dilution always leads to worse inference, as expected.

\paragraph{Convergence-breakdown for low seed probability}
A surprising behavior of the Belief Propagation equations was observed in \cite{ghio_bayes-optimal_2023}, for single instances at small values of $\gamma$, the patient zero probability. In fact, even in the Bayes optimal conditions, BP stops to converge. We checked that this breakdown of convergence is actually present even in the thermodynamic limit, using our population dynamic algorithm. This lack of convergence, therefore, seems to be related to a more profound reason. To understand what is happening, we simplified the framework by setting the infection probability $\lambda$ to 1 and by observing all the individuals at final time.
\begin{figure}
    \centering
    \includegraphics[width=0.45\textwidth]{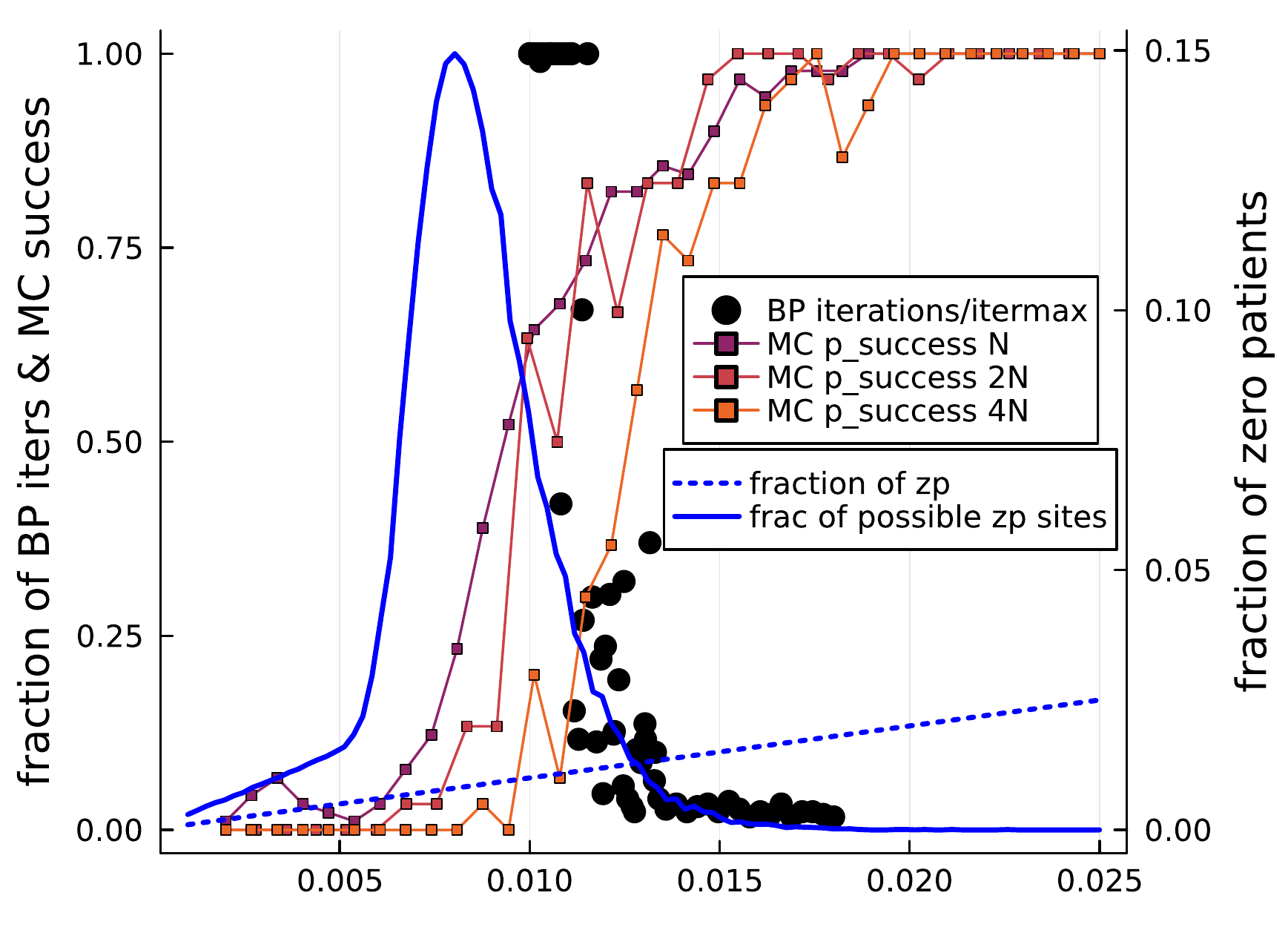}
    \caption{Study of population dynamics, Monte Carlo and number of clusters for zero-patience inference. We studied, for a RR graph with degree 3, the convergence of population dynamics (infinite graph) and Monte Carlo (finite-size graph). It breaks down at around $\tilde{\gamma}\simeq 0.013$. The black dots represent the number of iterations for population dynamics to reach convergence, normalized by the total number of iterations allowed. The continuous squared-marked lines represent the fraction of successful Monte Carlo runs. We say that MC is successful every time it reaches a configuration which satisfies the observations (see main text). The failure of Monte Carlo coincides with BP failure. We conjecture that this is due to Replica Symmetry Breaking. We think that the main reason of this breakdown is because the space in which a patient zero can be placed in the posterior becomes clustered (see Figure \ref{fig:hardshperes}). To support this conjecture, we plot the fraction of connected components of the sub-graph of all the individuals that could be the patients zero without violating the S observations. This number, as expected, grows sharply in the interval in which BP ceases to converge. The failure of convergence arises when the number of possible zones to place the patient zero (continue, blue line) becomes higher to the actual fraction of patients zero (dotted, blue line).  This suggests that when the number of zones in which a patient zero might be becomes larger than the number of patients zero, then the problem becomes hard, as illustrated in Figure \ref{fig:hardshperes}. }
    \label{fig:non_converg}
\end{figure}
In this regime, in Figure \ref{fig:non_converg} the black dots represent the number of iterations needed for the population dynamics algorithm to converge. Around $\gamma = \tilde{\gamma} = 0.013$ the algorithm stops converging. An intuitive explanation to explain this behaviour is the following: for $\gamma$ around $\tilde{\gamma}$ at final time many individuals are observed infectious (I) and a small (but extensive) part are observed susceptible (S). As $\lambda = 1$, the sole non-deterministic part of the process is the initial state, so the inference problem reduces to guess the position of the patients zero. The S individuals  not only signal that they were not infected in the epidemic process, but also that any patient zero must be at distance $>T$. For example, for a RR graph, this excludes a sphere centered in S-observed with $d(d-1)^{T-1}$ individuals. For $\gamma$ around $\tilde{\gamma}$ these spheres touch and intersect, so that the group of individuals eligible to be the patients zero gets separated in clusters.
\begin{figure}
    \centering
    \includegraphics[width=0.45\textwidth]{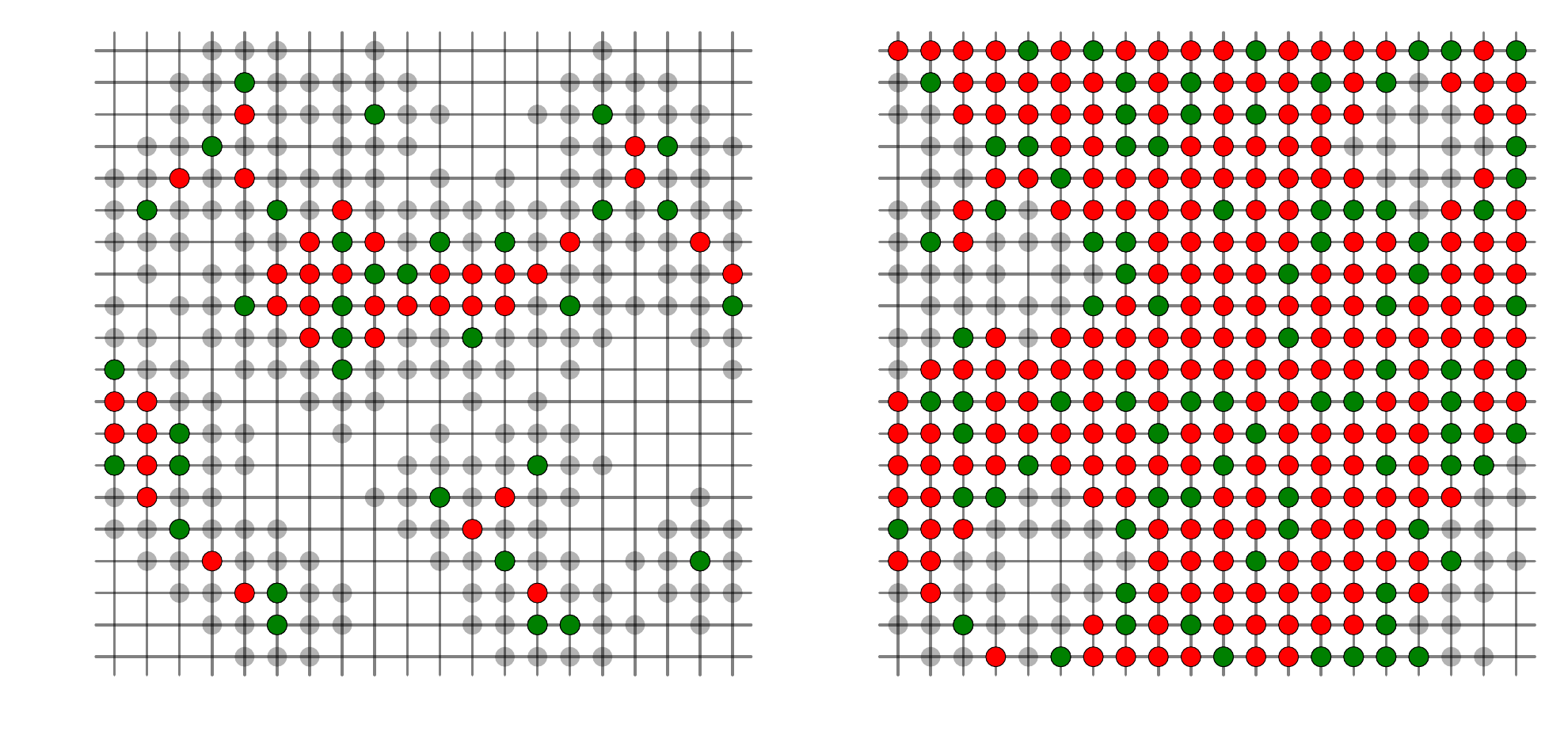}
    \caption{A 2D plot to visualize the geometric change undergone by the configuration space that could explain why population dynamics and Monte Carlo schemes stop to converge. In this plot, obtained by simulating epidemic spreading in a 2 dimensional lattice, we compare two scenarios. To the right, $\gamma$ is higher, namely there are more patients zero (green dots). This implies that the number of infected (greed, red and grey dots) is higher, so the number of S-observed individuals (no dots) is smaller. The patients zero can not be too close to the S-observed individuals because the infection probability is 1, so the observation constraint would be violated. The red dots represent all the individual which might be the patient zero according to the observations (i.e. individuals tested I and not too close to S-observed individuals). When the number of patients zero is lower, (\textit{left}) the number of S-observed individuals increases. So the possible zones to accommodate patients zero (green plus red dots) reduce and get clustered. This could create several separated states of the posterior, each one corresponding to a possible combination of placements of patients zero. }
    \label{fig:hardshperes}
\end{figure}
In Figure \ref{fig:hardshperes} we give a plot in 2D of the phenomenon. When the number of observed S is sufficiently high, due to the fact that there are few patients zero, the space in which patients zero could physically be gets fragmented. To check that this is what actually happens in a Random Regular graph, we initialized a graph and counted the connected components in which a patient zero could be present without violating the S observations. This number actually sharply increases in the decreasing $\gamma$ direction, around $\tilde{\gamma}$, i.e. when the algorithm stops converging. Further evidence of a phase transition is given by Monte-Carlo dynamics. We implemented a simple Monte Carlo simulation on a graph. We first sampled the planted (ground truth) configuration, from which we collected the observations. 
The observation protocol was set to observe all the individuals at the final time $T$ (without observation noise). Then we started a Metropolis-Hasting Monte Carlo simulation in order to sample a configuration satisfying all the observations. To do so, we initialize a configuration by doing a sample $x$ of the prior distribution. The initialization configuration typically does not satisfy the observations. So we make the following move: we randomly select an individual and we change its $t=0$ state by sampling the $I$ state with probability $\gamma$ (and the $S$ state with probability $(1-\gamma)$). Subsequently, the initial state configuration is evolved (deterministically, since the infection probability is $\lambda=1$). The configuration at final time is then checked to be consistent with the observations. In particular, we introduced the energy:
\begin{equation}
    U = -\sum_{i=1}^N\log p(o_i|x_i^T)
\end{equation}
Where $x$ is the configuration and each $o_i$ is the observation on the $i$th individual. In principle $p(o_i|x_i^T)$ should be either 0 (when the configuration does not satisfy the observations) or 1 (when the observation is satisfied).  In order to avoid infinite energy barriers, we introduced a small noise in observations, which we reduced during the Monte Carlo by means of an annealing procedure. In other words, the energy is just a penalization for each broken constraint. At each step, the move in the space of initial states described above  is made. The move is accepted by following a standard Metropolis scheme.
The MC stops when the configuration satisfies all constraints. For each value of $\gamma$ we repeated 60 times the MC scheme and computed the fraction of runs in which the algorithm was able to reach a configuration satisfying all observation constraints. We plot in Figure \ref{fig:non_converg}  the fraction of Monte Carlo processes that reached a configuration of the posterior. We clearly see that this quantity drops down around $\tilde{\gamma}$ . Due to the failure of BP equations (for finite and infinite graph), the explosion of possible patient zero zones and the failure of the Monte Carlo scheme we conjecture Replica Symmetry Breaking transition around $\tilde{\gamma}$.

\subsection{Departing from Bayes-optimal conditions.}
\label{subsec:results_outsideBO}
\begin{figure}
    \centering
    \includegraphics[width=0.4\textwidth]{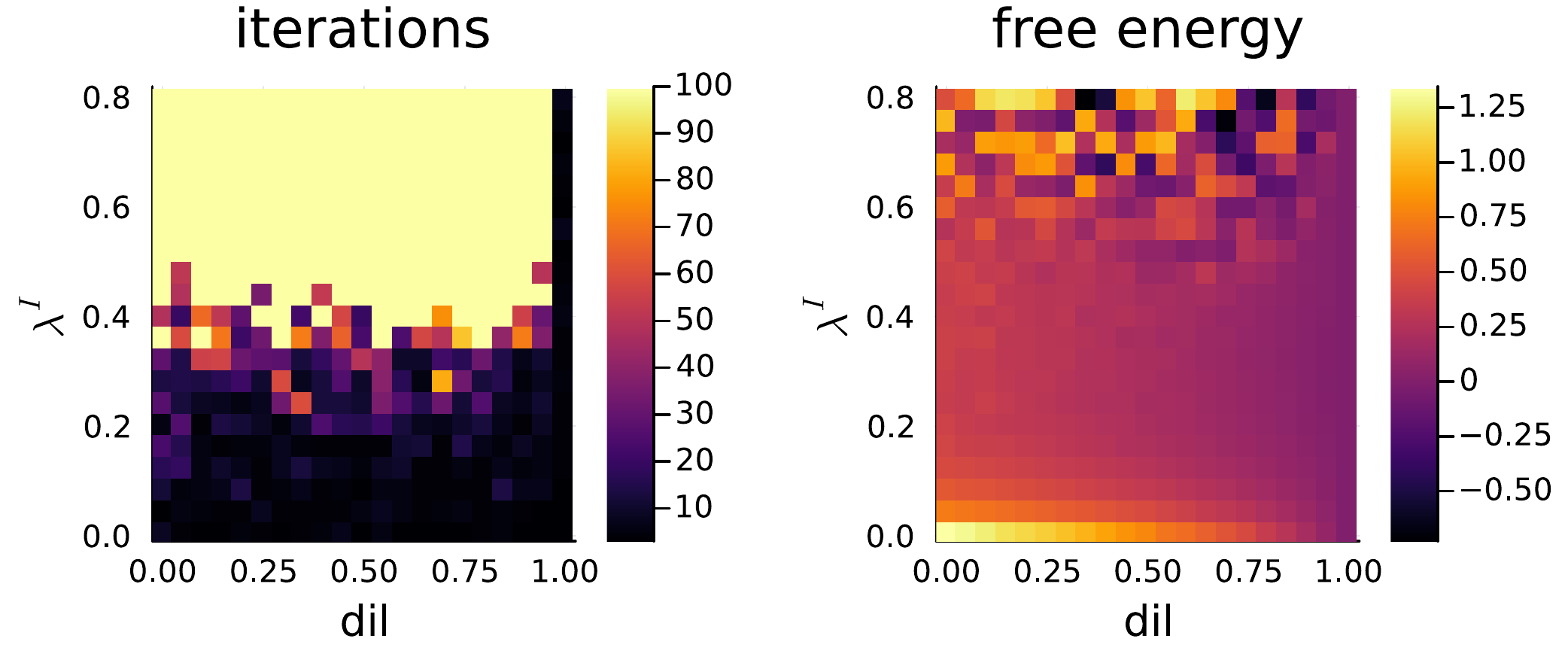}
    \caption{Ensemble algorithm breakdown outside Nishimori condition. Keeping fixed $\lambda^*=0.3$ and the parameters $\gamma^*=\gamma^I=0.03$, while instead moving the inference parameter $0.0<\lambda^I<0.8$ and the dilution between $0$ and $1$ we see that for high values of $\lambda^I$ the algorithm does not converge and provides nonphysical results for the free energy. }
    \label{fig:rsb}
\end{figure}
It is well known that when inference is performed with imperfect knowledge of the prior distribution parameters, it is possible to observe a Replica Symmetry Breaking (RSB) phase transition. This is due to the fact that outside the Bayes optimality regime, the Nishimori conditions are no longer guaranteed to be valid, see section~\ref{subsec:Nishimori}. An RSB phase can manifest itself with a convergence failure of the population dynamics algorithm, which is based on the Replica Symmetry hypothesis. This is exactly what we see in Figure \ref{fig:rsb}. We recall that we use a star ($^*$) to label the parameters with which the planted configuration is generated, e.g. $\lambda^*$ is the true infection probability, while $\lambda^I$ is the infection probability used by the algorithm in the inference process. 
For this plot we fixed $\gamma^*=\gamma^I$ (so we gave to the algorithm the exact value of patient zero probability) and we studied the free energy landscape by varying $\lambda^I$ and the observations dilution. There exists a zone in which the number of iterations reaches the maximum allowed number (which was set to $100$). In this zone, the observables show an oscillating behaviour. This suggests a breakdown of the algorithm validity, which may be caused by an RSB phase transition.   
In any case, when the prior is not known, some difficulties arise in epidemic inference. A good strategy to avoid them is to infer the prior parameters, as shown in the next paragraph.

\paragraph{Inferring epidemic prior's parameters.}
\begin{figure}
    \centering
    \includegraphics[width=0.35\textwidth]{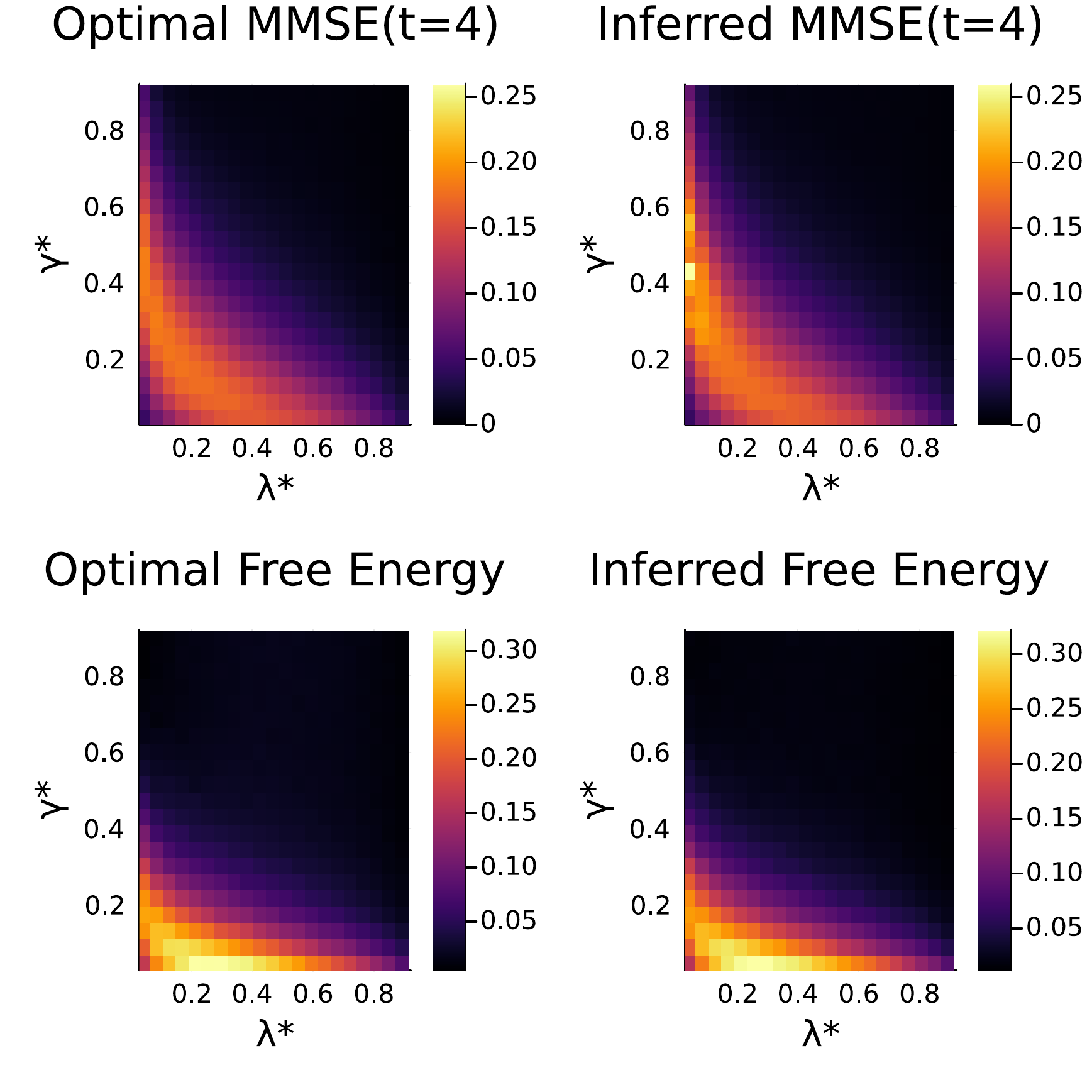}
    \caption{A comparison between the inference feasibility (quantified here with MMMSE and Free Energy) in the case in which prior's parameters are known (\textit{first column}) and when instead they are learnt (\textit{second column}). The quantities are represented as functions of the planted parameters $\gamma^*$ and $\lambda^*$. In the first row the MMSE at intermediate time $t=4$ is plotted: on the left there is the Optimal Bayes result, already shown in Figure \ref{fig:comp_measures}, while on the right there is the result obtained when the infection and patient zero parameters $\lambda^I$ and  $\gamma^I$ are learnt. On the second row the same comparison (i.e. Bayes optimality on the left and hyper-parameters' learning on the right) is done for free energy. In both cases (MMSE and Free Energy) the initial conditions for the hyperparameters were set to $\lambda^I=0.5$ and $\gamma^I=0.5$. The results are for the Erd\"os-R\'enyi  ensemble with average degree of $3$. Observations are made at final time $T=8$.}
    \label{fig:gamVSlamLEARN}
\end{figure}
We infer the prior parameters by (approximately) minimizing the Bethe Free Energy. In particular, for the patient zero probability $\gamma$ we use the Expectation Maximization (EM) method. For the infection probability $\lambda$, instead, we perform a gradient descent (GD) on the free energy. This mixed strategy (EM for inferring $\gamma$ and GD for $\lambda$) was adopted due to its simplicity in terms of calculations. To check if the method works, we studied inference in the same conditions of Figure \ref{fig:comp_measures}. We therefore fixed observations dilution to $0.5$, and we explored the space of patient zero probability $\gamma^*$ against infection probability $\lambda^*$.
Initializing the inferring parameters to $\gamma^I=\lambda^I=0.5$, the results are shown in Figure \ref{fig:gamVSlamLEARN}. The plot shows a comparison between the observables computed by inferring the prior parameters and their respective quantities in the Bayes optimal case, i.e. the ones plotted in Figure \ref{fig:comp_measures} (first row) and Figure \ref{fig:free_energies} (left).
The prior parameters are learnt by minimizing the free energy, which agrees almost perfectly with the optimal one. There is a strong agreement also for other observables, as the MMSE, which we plotted at time $t=4$
\begin{figure}[t]
    \centering
    \includegraphics[width=0.4\textwidth]{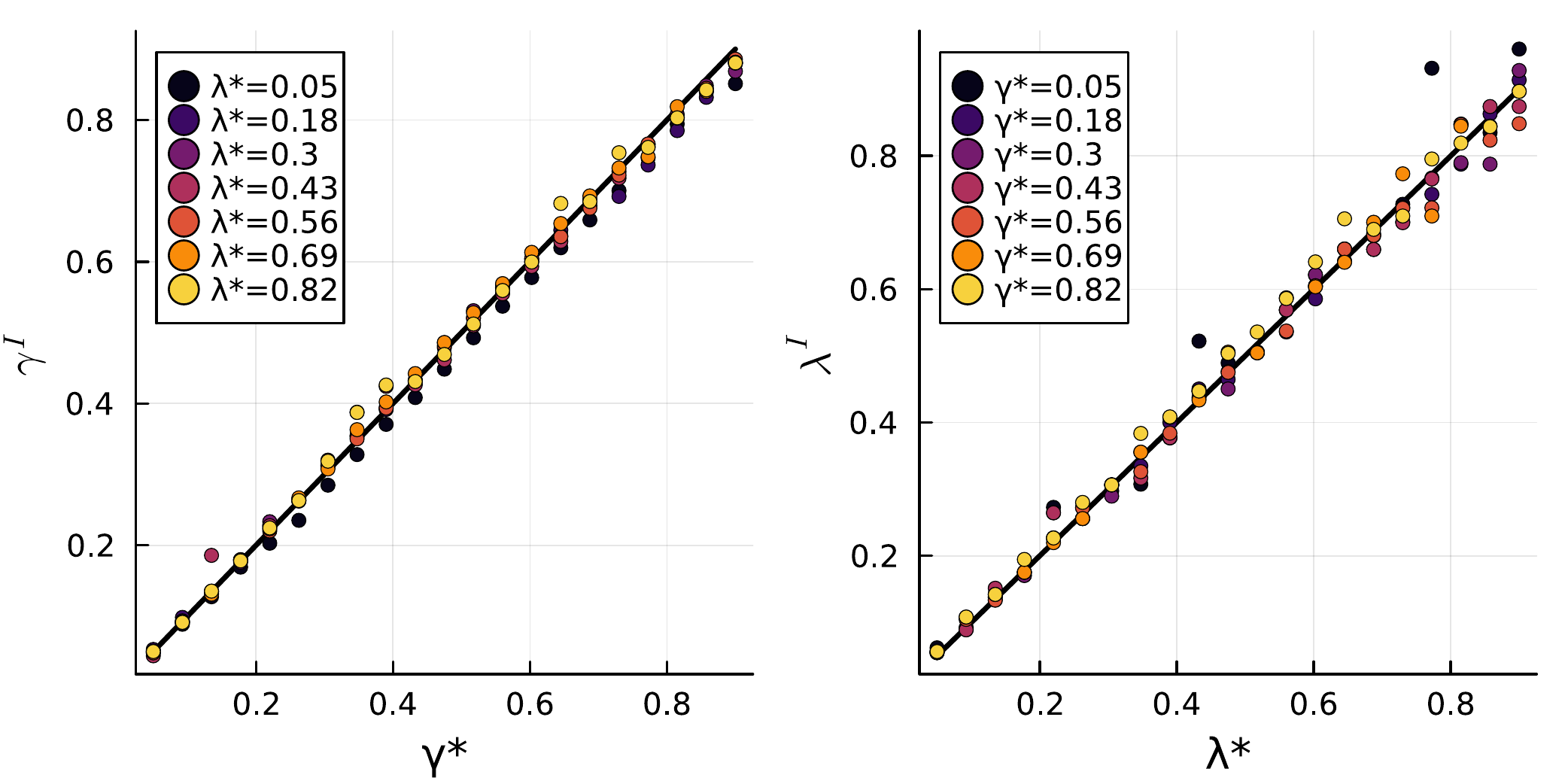}
    \caption{The inferred prior parameters in function of their respective planted quantities. The plot is obtained at zero dilution (all individual observed) and for (uniformly) scattered observations in time. 
    In the left panel, patient zero parameter $\gamma^I$ is plotted in function of  $\gamma^*$ for different values of $\lambda^*$. The right panel's lines are instead the values of the infection $\lambda^I$ as function of $\lambda^*$ for different values of $\gamma^*$.}
    \label{fig:prior_params}
\end{figure}
To actually see how well the prior hyper-parameters are inferred, we plot them in function of their planted respective quantity (see Figure~\ref{fig:prior_params}). 
\begin{figure}
    \centering
    \includegraphics[width=0.35\textwidth]{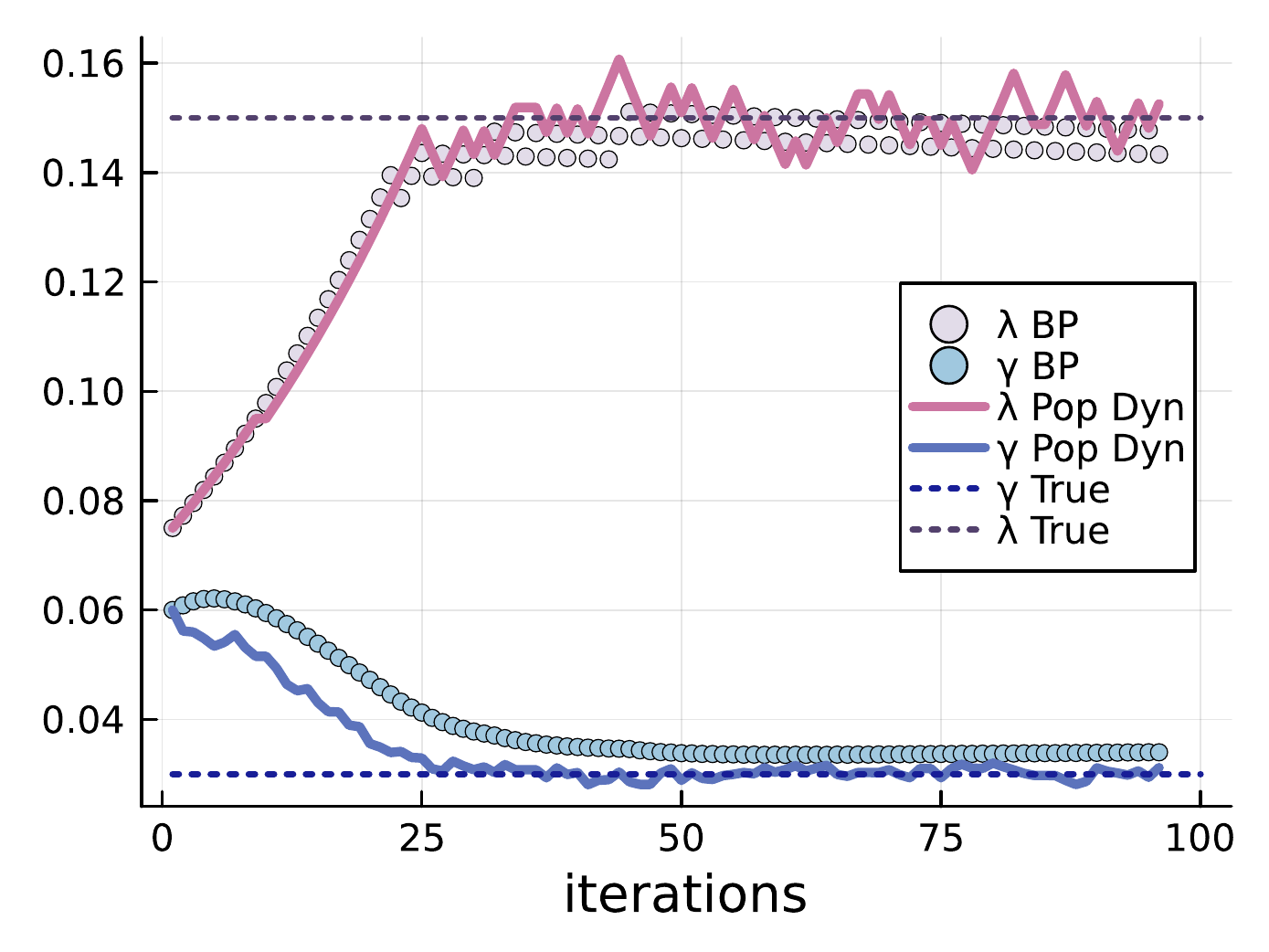}
    \caption{Inference of prior parameters with the ensemble code (Pop Dynamics) compared to the single instance result, obtained running the Belief Propagation (BP) algorithm on a contact network of $N=10000$ nodes. The plot shows the gradient descent in free energy with respect to the two parameters $\gamma^I$ and $\lambda^I$ which respectively represent the patient zero and the infection probabilities. For the infection $\lambda^I$ the gradient descent is performed using the Sign Descender technique with learning rate $0.01$, while for $\gamma^I$ we used the Expectation Maximization method. The results are for Erd\"os-R\'enyi (ER) graphs with average degree $3$. All the individuals are observed at final time.}
    \label{fig:hyperparams_sib_ens}
\end{figure}
It is again important to compare the results of prior parameters inference with the single instance results on finite graphs. Indeed, the inference results shown so far are for infinitely large graphs. The number of observations is therefore infinite too. It is then crucial to see whether for finite size graphs (and finite information) it is possible to achieve comparable results to the ensemble. In Figure \ref{fig:hyperparams_sib_ens} we see that this is the case. We compare the population dynamics and the single instance code by analyzing step-by-step their gradient descent on the patient zero and infection probabilities. The plot shows that, as expected, the ensemble inference is more precise due to infinite amount of information available. However, the values inferred by the single instance algorithm are very close to the true ones.

\paragraph{Addressing biased observations}

In realistic contexts, observations are not taken uniformly at random
from the population. This is because infected people might manifest
some symptoms, which push them to test themselves. The probability
of being observed is therefore typically higher for infected people
than for susceptible ones. A first consequence is that the fraction of
infected individuals in the population is not equal to the fraction of infected ones in the set of observed
individuals. If in the inference process this is not considered, the
risk is to achieve low performance. Non considering the bias of observations
means to infer with an incorrect prior parameter, i.e. outside of
the Nishimori conditions. 
To quantify this bias, we introduce $p_+$, the probability for an infected individual to be symptomatic.
We assume that all infected symptomatic individuals are tested.
Asymptomatic individuals are instead tested at random with probability $p_{r}$. 
From this, the probability for an infected individual to be tested, with a positive test result is: 
\begin{align*}
P({\rm tested, positive}|I) =(1-\text{fr})(p_{+}+p_{r}(1-p_{+}))
\end{align*}
and similarly:
\[
P({\rm tested, negative}|I) =\text{fr}(p_{+}+p_{r}(1-p_{+}))
\]
For susceptible states $S$:
\begin{align*}
P({\rm tested, positive}|S) & =p_{r}\text{fr}\\
P({\rm tested, negative}|S) & =p_{r}(1-\text{fr})
\end{align*}
The unbiased case is recovered for $p_{+}=0$.
\begin{figure}
    \centering
    \includegraphics[width=0.5\textwidth]{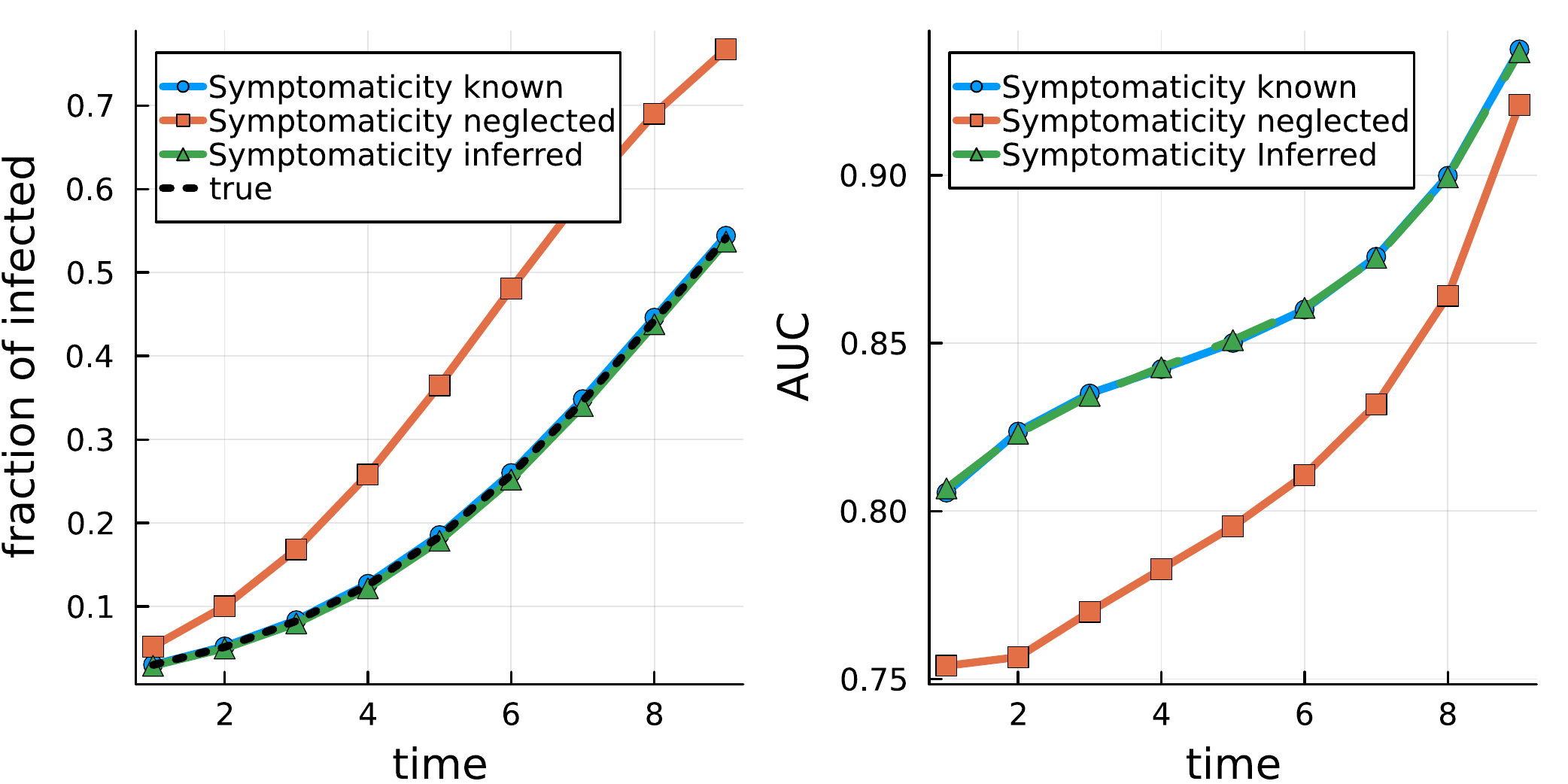}
    \caption{Considering (and inferring) bias in observation allows to recover Nishimori conditions and improves inference performance. The bias is generated by symptomatic individuals, which are all assumed to be tested. The probability for an infected individual of being symptomatic was set to $p_+ = 0.5$. Asymptomatic individuals can also be tested. For this plot, the probability for an asymptomatic individual to be randomly selected for a test was set to $p_r = 0.04$. The left plot shows the estimated fraction of infected individuals over time. Considering the bias in the inference process allows to reconstruct this function. On the right plot we compared inference performance when the bias is considered VS when it is neglected. Considering the bias systematically improved the AUC. The patient zero probability was set to $\gamma=0.03$ and the infection probability to $\lambda=0.25$. The observations are all performed at time $T=8$.}
    \label{fig:osb_trig}
\end{figure}
We want to compare inference results when the bias $p_+$ is considered and when instead is neglected. In Figure \ref{fig:osb_trig} (left panel), we see a substantial overestimation of the infection when ignoring the bias. In the  right panel, we see that the AUC is systematically higher when the bias is included. We finally inferred the bias $p_+$ by minimizing the Free Energy (following exactly the same procedure of the transmission rate's inference). This process allows to include the unknown bias without affecting performances.

\section{Conclusion}
In this paper, we study the feasibility of inference in epidemic spreading on a contact graph. 
Using the Replica Symmetric cavity method, we give quantitative predictions of several estimators (Minimum Mean Square Error MMSE, Maximum Mean Overlap MMO, and Area Under the ROC curve AUC), in different regimes depending on the characteristics of the epidemics, the observations and the contact graph.

In the Bayes-optimal setting, we show that for a large range of the model's parameters, the RS predictions are in good agreement with the results obtained on finite size instances.
It was noted in \cite{ghio_bayes-optimal_2023} that BP equations did not converge on large instances in a particular region of the parameters  (at low seed probability and high rate transmission), a fact that is also confirmed by our simulations.
Our simulations in that region show a lack of convergence of the cavity equations in the thermodynamic limit (answering thus negatively to the conjecture in  \cite{ghio_bayes-optimal_2023} of it being to finite-size effects), and strongly hinting to Replica Symmetry Breaking.

In the non-Bayes optimal setting (i.e. when the parameters of the posterior differ from the parameters used in the prior), we observe a region where the iterative numerical resolution of the Replica Symmetric cavity equations does not converge, suggesting the presence of a Replica Symmetry Breaking transition.
We show however that inferring parameters allowing to recover performance comparable to the one obtained in the Bayes optimal setting is possible with a simple iterative procedure, for a large range of the prior's parameters. There are however situations in which one is forced to work outside the Bayes optimal case (and for which Replica Symmetry Breaking is to be expected): e.g. when some parameters of the model are known only approximately but are too many to infer, or when the knowledge of the contact network itself is imperfect.

Averaging over correlated disorder within the framework of the cavity method is the main technical issue addressed in this paper. 
The strategy developed here could be applied to more involved irreversible epidemic models, such as the SIR and SEIR model. 
The main limitation would be an increase of space size of the dynamical variables: each compartment added would come with a additional couple of transition times (one planted and one inferred time). 
The strategy could be applied more generally to any model in which disorder can be decomposed into a set of local (independent) random variables $\underline{s}$, and a set of correlated variables $\underline{\tau}$ that can be computed from the first set.
Note however that each element of the correlated disorder $\underline{\tau}$ should be expressed only as a  function of a local subset of  $\underline{\tau}$ and $\underline{s}$. In other words, there must exist a function:
\begin{equation}
    \psi(\underline{\tau}|\underline{s}) = \prod_{i\in V}\psi_i(\tau_i|\underline{\tau}_{\partial i},\underline{s}_{\partial i})
\end{equation}
with arbitrary factors $\psi_i$, which for fixed $\underline{s}$ is non-zero only for a given value $\underline{\tau}$. 
\section{Acknowledgement}
This study was carried out within the FAIR - Future Artificial Intelligence Research and received funding
from the European Union Next-GenerationEU (Piano Nazionale Di Ripresa e Resilienza (PNRR) – Missione
4 Componente 2, Investimento 1.3 – D.D. 1555 11/10/2022, PE00000013). This manuscript reflects only the
authors’ views and opinions, neither the European Union nor the European Commission can be considered
responsible for them.
\bibliography{draft}

\onecolumngrid
\newpage
\appendix
\section{BP equations and Bethe Free Energy}
\label{app:BP_derivation}

\begin{figure}
    \centering
    \includegraphics[width=0.7\textwidth]{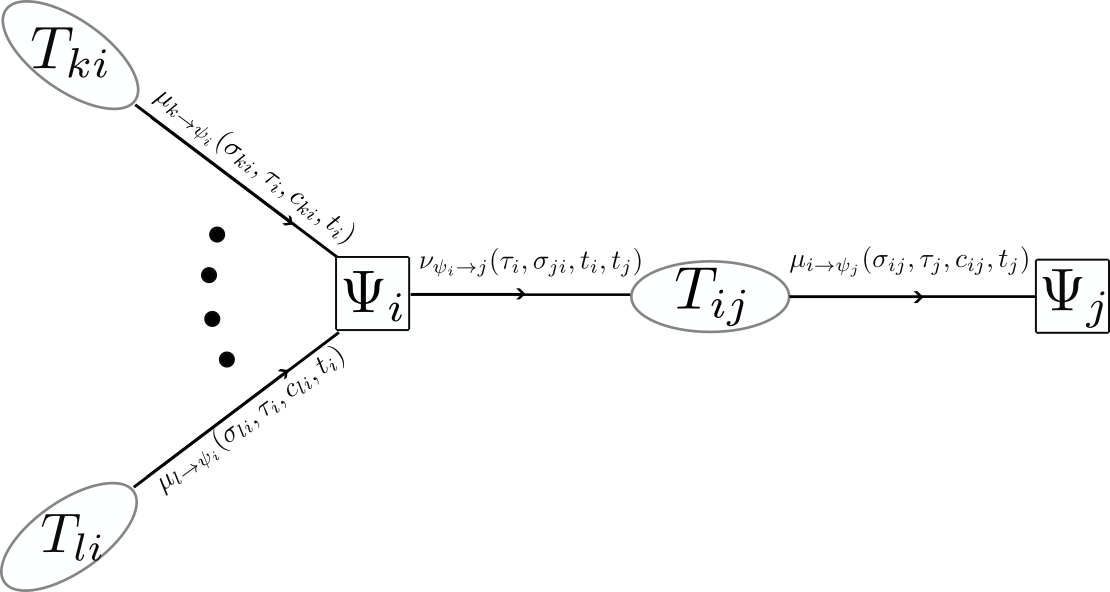}
    \caption{The compact optimized version of BP equations. The $\mu$ messages are functions of $\propto T^2$ values, since $\sigma \in \{0,1,2\}$ and $c\in\{0,1\}$. For this reason we keep the population of $\mu$ messages. Each iteration of the optimized BP consists in computing the $\nu$  message from a set of $\mu$ messages and after the extraction of disorder, according to equation \eqref{eq:BP_factor_to_variable}. Then from the $\nu$  message, by performing the summation on the last argument described in equation \eqref{eq:BP_variable_to_factor}, the new $\mu$ message is obtained. }
    \label{fig:messages}
\end{figure}

In this appendix we derive a simplified version of the BP equations (\ref{eq:BP_equations}) introduced in section \ref{subsec:BP_for_joint}. These simplified equations (given in (\ref{eq:BP_factor_to_variable}) and (\ref{eq:BP_variable_to_factor})) are over a set of modified messages represented in Figure \ref{fig:messages}.
\subsection{Clamping} In the numerical resolution of the cavity equations, it will be convenient to introduce a horizon time $T+1$ above which the epidemic evolution is not observed.
This results in a modification of the function $\psi^*$ ensuring the constraints on infection times:
\begin{align}
	\psi^*(\tau_i,\underline{\tau}_{\partial i},x_i^0,\{s_{ji}\}) &= \mathbb{I}[\tau_i=\delta_{x_i^0,S}\min(T+1,\min_{l\in\partial i}(\tau_l+s_{li}))] \ .
\end{align}
\subsection{Simplifications}

In order to simplify the BP equations~\ref{eq:BP_equations}, we will start by writing the functions $\psi^*,\psi$ in a simplified way:
\begin{align}
	\psi^*(\tau_i^{(j)}, \underline{\tau}_{\partial i}^{(i)},\{s_{li}\}_{l\in\partial i},x_i^0) &= \delta_{x_i^0,I}\delta_{\tau_i^{(j)},0} + \delta_{x_i^0,S}\prod_{l\in\partial i}\mathbb{I}[\tau_i^{(j)}\leq\tau_l^{(i)}+s_{li}] - \delta_{x_i^0,S}\mathbb{I}[\tau_i^{(j)}<T+1]\prod_{l\in\partial i}\mathbb{I}[\tau_i^{(j)}<\tau_l^{(i)}+s_{li}]
\end{align}
and:
\begin{align}
\begin{aligned}
	\psi(t_i^{(j)}, \underline{t}_{\partial i}^{(i)}) &= \sum_{x_i^0}\gamma(x_i^0)\sum_{\{s_{li}\}_{l\in\partial i}}\prod_{l\in\partial i}w(s_{li})\mathbb{I}[t_i^{(j)}=\delta_{x_i^0,S}\min(T+1,t_l^{(i)}+s_{li})] \\
	&= \gamma \delta_{\tau_i^{(j)},0} + (1-\gamma)\left[\prod_{l\in\partial_i}\left(\sum_{s=1}^{\infty}w(s)\mathbb{I}[t_i^{(j)}\leq t_l^{(i)}+s]\right) - \mathbb{I}[\tau_i^{(j)}<T+1]\prod_{l\in\partial_i}\left(\sum_{s=1}^{\infty}w(s)\mathbb{I}[t_i^{(j)}< t_l^{(i)}+s]\right)\right]\\
	&= \gamma \delta_{\tau_i^{(j)},0} + (1-\gamma)\left[\prod_{l\in\partial_i}a(t_i^{(j)}-t_l^{(i)}-1) - \mathbb{I}[\tau_i^{(j)}<T+1]\prod_{l\in\partial_i}a(t_i^{(j)}-t_l^{(i)})\right]\\
	&= \gamma(t_i^{(j)})\left(\prod_{l\in\partial i} a(t_i^{(j)}-t_l^{(i)}-1) - \phi(t_i^{(j)})\prod_{l\in\partial i}a(t_i^{(j)}-t_l^{(i)}) \right)
\end{aligned}
\end{align}
where we have defined:
\begin{align}
\begin{aligned}
	a(t) &= (1-\lambda)^{tH(t)} \\
	\gamma(t) &= \begin{cases}
	\gamma & \text{if} \quad t=0\\
	1-\gamma & \text{if} \quad t>0
	\end{cases} \ .\\
	\phi(t) &= \begin{cases}
	0 & \text{if} \quad t=0 \ \text{or} \ t=T+1\\
	1 & \text{if} \quad 0<t<T+1
	\end{cases} \ .
\end{aligned}	
\end{align}
where $H(t)$ is the Heaviside step function, with $H(0)=0$.
We also notice that the function $\Psi$ constraints the planted and inferred times of the incoming messages to the equality: $\tau_i^{(k)}=\tau_i^{(j)}$, and $t_i^{(k)}=t_i^{(j)}$ for all $k\in\partial i \setminus j$.
We can now re-write the first BP equation with the expression of $\psi^*, \psi$:
\begin{align}
\begin{aligned}
	\nu_{\Psi_i\to j}(T_{ij}) &=\frac{\gamma(t_i^{(j)})\xi(\tau_i^{(j)},t_i^{(j)},c_i)}{z_{\Psi_i\to j}}\left(
	a(t_i^{(j)}-t_j^{(i)}-1)\delta_{x_i^0,I}\delta_{\tau_i^{(j)},0}\prod_{k\in\partial i\setminus j}\left[\sum_{t_k^{(i)}}a(t_i^{(j)}-t_k^{(i)}-1)\sum_{\tau_k^{(i)}}\mu_{k\to \Psi_i}(T_{ki})\right]\right.\\
	+&a(t_i^{(j)}-t_j^{(i)}-1)\delta_{x_i^0,S}\mathbb{I}[\tau_i^{(j)}\leq\tau_j^{(j)}+s_{ji}]\prod_{k\in\partial i\setminus j}\left[\sum_{t_k^{(i)}}a(t_i^{(j)}-t_k^{(i)}-1)\sum_{\tau_k^{(i)}}\mu_{k\to \Psi_i}(T_{ki})\mathbb{I}[\tau_i^{(j)}\leq\tau_k^{(i)}+s_{ki}]\right]\\
	-&a(t_i^{(j)}-t_j^{(i)}-1)\delta_{x_i^0,S}\mathbb{I}[\tau_i^{(j)}<T+1]\mathbb{I}[\tau_i^{(j)}<\tau_j^{(j)}+s_{ji}]\\
	 &\qquad\qquad\qquad\times\prod_{k\in\partial i\setminus j}\left[\sum_{t_k^{(i)}}a(t_i^{(j)}-t_k^{(i)}-1)\sum_{\tau_k^{(i)}}\mu_{k\to \Psi_i}(T_{ki})\mathbb{I}[\tau_i^{(j)}<\tau_k^{(i)}+s_{ki}]\right]\\
	-&\phi(t_i^{(j)})a(t_i^{(j)}-t_j^{(i)})\delta_{x_i^0,I}\delta_{\tau_i^{(j)},0}\prod_{k\in\partial i\setminus j}\left[\sum_{t_k^{(i)}}a(t_i^{(j)}-t_k^{(i)})\sum_{\tau_k^{(i)}}\mu_{k\to \Psi_i}(T_{ki})\right]\\
	-&\phi(t_i^{(j)})a(t_i^{(j)}-t_j^{(i)})\delta_{x_i^0,S}\mathbb{I}[\tau_i^{(j)}\leq\tau_j^{(j)}+s_{ji}]\prod_{k\in\partial i\setminus j}\left[\sum_{t_k^{(i)}}a(t_i^{(j)}-t_k^{(i)})\sum_{\tau_k^{(i)}}\mu_{k\to \Psi_i}(T_{ki})\mathbb{I}[\tau_i^{(j)}\leq\tau_k^{(i)}+s_{ki}]\right]\\
	+&\phi(t_i^{(j)})a(t_i^{(j)}-t_j^{(i)})\delta_{x_i^0,S}\mathbb{I}[\tau_i^{(j)}<T+1]\mathbb{I}[\tau_i^{(j)}<\tau_j^{(j)}+s_{ji}]\\
	&\left.\qquad\qquad\qquad\times\prod_{k\in\partial i\setminus j}\left[\sum_{t_k^{(i)}}a(t_i^{(j)}-t_k^{(i)})\sum_{\tau_k^{(i)}}\mu_{k\to \Psi_i}(T_{ki})\mathbb{I}[\tau_i^{(j)}<\tau_k^{(i)}+s_{ki}]\right]\right)
\end{aligned}
\end{align}
where $T_{ki}=(\tau_k^{(i)},\tau_i^{(k)}=\tau_i^{(j)},t_k^{(i)},t_i^{(k)}=t_i^{(j)})$ in the r.h.s., due to the constraint on the incoming times (and $T_{ij} = (\tau_i^{(j)},\tau_j^{(i)},t_i^{(j)},t_j^{(i)})$ in the l.h.s.). 

\subsection{Summation over the planted times}
We can see on the above equation that the r.h.s. depends on the planted time $\tau_j^{(i)}$ only through the sign:
\begin{align}
\label{eq:def_sigma_ji}
	\sigma_{ji} = 1+\text{sgn}(\tau_j^{(i)}-\tau_i^{(j)}+s_{ji})\,
\end{align}
with the convention that $\text{sgn}(0)=0$.
We therefore introduce the notation	:
\begin{align}
\label{eq:tilde_nu}
	\tilde{\nu}_{\Psi_i\to j}(\tau_i^{(j)}, \sigma_{ji},t_i^{(j)},t_j^{(i)})=\nu_{\Psi_i\to j}(\tau_i^{(j)},\tau_j^{(i)},t_i^{(j)},t_j^{(i)})
\end{align}
for all $\tau_j^{(i)}$ such that $\sigma_{ji} = 1+\text{sgn}(\tau_j^{(i)}-\tau_i^{(j)}+s_{ji})$.
We also introduce the message:
\begin{align}
	\tilde{\mu}_{i\to \Psi_j}(\sigma_{ij},\tau_j^{(i)},t_i^{(j)},t_j^{(i)}) = \sum_{\tau_i^{(j)}}\mu_{i\to \Psi_j}(\tau_i^{(j)},\tau_j^{(i)},t_i^{(j)},t_j^{(i)})\mathbb{I}[\sigma_{ij}=1+\text{sgn}(\tau_i^{(j)}-\tau_j^{(i)}+s_{ij})]
\end{align}
With these definitions, the BP equation becomes:
\begin{align}
\begin{aligned}
	\tilde{\nu}_{\Psi_i\to j}(\widetilde{T}_{ij}) &=\gamma(t_i^{(j)})\xi(\tau_i^{(j)},t_i^{(j)},c_i)\left( a(t_i^{(j)}-t_j^{(i)}-1)\delta_{x_i^0,I}\delta_{\tau_i^{(j)},0}\prod_{k\in\partial i\setminus j}\left[\sum_{t_k^{(i)}}a(t_i^{(j)}-t_k^{(i)}-1)\sum_{\sigma_{ki}=0}^2\tilde{\mu}_{k\to \Psi_i}(\widetilde{T}_{ki})\right]\right.\\
	+&a(t_i^{(j)}-t_j^{(i)}-1)\delta_{x_i^0,S}\mathbb{I}[\sigma_{ji}\in\{1,2\}]\prod_{k\in\partial i\setminus j}\left[\sum_{t_k^{(i)}}a(t_i^{(j)}-t_k^{(i)}-1)\sum_{\sigma_{ki}=1}^2\tilde{\mu}_{k\to \Psi_i}(\widetilde{T}_{ki})\right]\\
	-&a(t_i^{(j)}-t_j^{(i)}-1)\delta_{x_i^0,S}\mathbb{I}[\tau_i^{(j)}<T+1]\mathbb{I}[\sigma_{ji}=2]\prod_{k\in\partial i\setminus j}\left[\sum_{t_k^{(i)}}a(t_i^{(j)}-t_k^{(i)}-1)\tilde{\mu}_{k\to \Psi_i}(\sigma_{ki}=2,\tau_i^{(j)},t_k^{(i)},t_i^{(j)})\right]\\
	-&\phi(t_i^{(j)})a(t_i^{(j)}-t_j^{(i)})\delta_{x_i^0,I}\delta_{\tau_i^{(j)},0}\prod_{k\in\partial i\setminus j}\left[\sum_{t_k^{(i)}}a(t_i^{(j)}-t_k^{(i)})\sum_{\sigma_{ki}=0}^2\tilde{\mu}_{k\to \Psi_i}(\widetilde{T}_{ki})\right]\\
	-&\phi(t_i^{(j)})a(t_i^{(j)}-t_j^{(i)})\delta_{x_i^0,S}\mathbb{I}[\sigma_{ji}\in\{1,2\}]\prod_{k\in\partial i\setminus j}\left[\sum_{t_k^{(i)}}a(t_i^{(j)}-t_k^{(i)})\sum_{\sigma_{ki}=1}^2\tilde{\mu}_{k\to \Psi_i}(\widetilde{T}_{ki})\right]\\
	+&\left.\phi(t_i^{(j)})a(t_i^{(j)}-t_j^{(i)})\delta_{x_i^0,S}\mathbb{I}[\tau_i^{(j)}<T+1]\mathbb{I}[\sigma_{ji}=2]\prod_{k\in\partial i\setminus j}\left[\sum_{t_k^{(i)}}a(t_i^{(j)}-t_k^{(i)})\tilde{\mu}_{k\to \Psi_i}(\sigma_{ki}=2,\tau_i^{(j)},t_k^{(i)},t_i^{(j)})\right]\right)
\end{aligned}
\end{align}
where $\widetilde{T}_{ij} = (\tau_i^{(j)},\sigma_{ji},t_i^{(j)},t_j^{(i)})$, and $\widetilde{T}_{ki}=(\sigma_{ki}, \tau_i^{(k)}=\tau_i^{(j)}, t_k^{(i)},t_i^{(k)}=t_i^{(j)})$ for all $k\in\partial i \setminus j$.
In the above equation we have dropped the normalization factor $z_{\Psi_i\to j}$, since the message $\tilde{\nu}_{\Psi_i\to j}$ is not a probability but the value taken by the (normalized) BP message $\nu_{\Psi_i\to j}$ for any $\tau_j^{(i)}$ achieving the equality (\ref{eq:def_sigma_ji}).
The other BP equation becomes:
\begin{align}
\begin{aligned}
	\tilde{\mu}_{i\to \Psi_j}(\sigma_{ij},\tau_j^{(i)},t_i^{(j)},t_j^{(i)}) &= \sum_{\tau_i^{(j)}=0}^{{T+1}}\mu_{i\to \Psi_j}(\tau_i^{(j)},\tau_j^{(i)},t_i^{(j)},t_j^{(i)})\mathbb{I}[\sigma_{ij}=1+\text{sgn}(\tau_i^{(j)}-\tau_j^{(i)}+s_{ij})] \\
	&= \sum_{\tau_i^{(j)}=0}^{{T+1}}\nu_{\Psi_i\to j}(\tau_i^{(j)},\tau_j^{(i)},t_i^{(j)},t_j^{(i)})\mathbb{I}[\sigma_{ij}=1+\text{sgn}(\tau_i^{(j)}-\tau_j^{(i)}+s_{ij})] \\
	&= \sum_{\tau_i^{(j)}=0}^{{T+1}}\tilde{\nu}_{\Psi_i\to j}(\tau_i^{(j)},\sigma_{ji}=1+\text{sgn}(\tau_j^{(i)}-\tau_i^{j}+s_{ji}),t_i^{(j)},t_j^{(i)})\mathbb{I}[\sigma_{ij}=1+\text{sgn}(\tau_i^{(j)}-\tau_j^{(i)}+s_{ij})]
\end{aligned}
\end{align}
which gives for each value of $\sigma_{ij}$:
\begin{align}
\begin{aligned}
\left\{
\begin{array}{llllll}
	\tilde{\mu}_{i\to \Psi_j}(0,\tau_j^{(i)},t_i^{(j)},t_j^{(i)})&=\mathbb{I}[\tau_j-s_{ji}>0]\sum_{\tau_i^{(j)}=0}^{\tau_j^{(i)}-s_{ji}}\tilde{\nu}_{\Psi_i\to j}(\tau_i^{(j)},\sigma_{ji}=2,t_i^{(j)},t_j^{(i)}) \\
	\tilde{\mu}_{i\to \Psi_j}(1,\tau_j^{(i)},t_i^{(j)},t_j^{(i)})&= \mathbb{I}[\tau_j-s_{ji}\geq 0]\tilde{\nu}_{\Psi_i\to j}(\tau_i^{(j)}=\tau_j^{(i)}-s_{ji},\sigma_{ji}=2,t_i^{(j)},t_j^{(i)})\\	
	\tilde{\mu}_{i\to \Psi_j}(2,\tau_j^{(i)},t_i^{(j)},t_j^{(i)})&= \sum_{\tau_i^{(j)}=\zeta_{ij}^+}^{{T+1}}\tilde{\nu}_{\Psi_i\to j}(\tau_i^{(j)},\sigma_{ji}=1+\text{sgn}(\tau_j^{(i)}-\tau_i^{j}+s_{ji}),t_i^{(j)},t_j^{(i)})\\
	&=\sum_{\tau_i^{(j)}=\zeta^+_i}^{\zeta^-_i}\tilde{\nu}_{\Psi_i\to j}(\tau_i^{(j)},\sigma_{ji}=2,t_i^{(j)},t_j^{(i)}) \\
	&+ \mathbb{I}[\tau_j^{(i)}+s_{ji}\leq {T+1}]\tilde{\nu}_{\Psi_i\to j}(\tau_i^{(j)}=\tau_j^{(i)}+s_{ji},\sigma_{ji}=1,t_i^{(j)},t_j^{(i)})\\
	&+ \mathbb{I}[\tau_j^{(i)}+s_{ji}<{T+1}]\sum_{\tau_i^{(j)}=\tau_j^{(i)}+s_{ji}+1}^{T+1} \tilde{\nu}_{\Psi_i\to j}(\tau_i^{(j)},\sigma_{ji}=0,t_i^{(j)},t_j^{(i)})
\end{array}
\right.
\end{aligned}
\end{align}
where $\zeta^+_i=\max(0,\tau_j^{(i)}-s_{ij}+1)$, and $\zeta^-_i=\min(T+1,\tau_j^{(i)}+s_{ji}-1)$.

\subsection{Summation over the inferred times}
In order to reduce further the space of variables over which the BP messages are defined, we define the following message:
\begin{align}
	\mu'_{i\to\Psi_j}(\sigma_{ij},\tau_j^{(i)},c_{ij},t_j^{(i)})=\sum_{t_i^{(j)}}\tilde{\mu}_{i\to\Psi_j}(\sigma_{ij},\tau_j^{(i)},t_i^{(j)},t_j^{(i)})a(t_j^{(i)}-t_i^{(j)}-c_{ij}) \ ,
\end{align}
with $c_{ij}\in\{0,1\}$.
Using this definition, the first BP equation becomes:
\begin{align}
\label{eq:BP_factor_to_variable}
\begin{aligned}
	\tilde{\nu}_{\Psi_i\to j}(\tau_i^{(j)},\sigma_{ji},t_i^{(j)},t_j^{(i)}) &=\gamma(t_i^{(j)})\xi(\tau_i^{(j)},t_i^{(j)},c_i)\left( a(t_i^{(j)}-t_j^{(i)}-1)\delta_{x_i^0,I}\delta_{\tau_i^{(j)},0}\prod_{k\in\partial i\setminus j}\left[\sum_{\sigma_{ki}=0}^2 \mu'_{k\to \Psi_i}(\sigma_{ki},\tau_i^{(k)},c_{ki}=1,t_i^{(k)})\right]\right.\\
	+&a(t_i^{(j)}-t_j^{(i)}-1)\delta_{x_i^0,S}\mathbb{I}[\sigma_{ji}\in\{1,2\}]\prod_{k\in\partial i\setminus j}\left[\sum_{\sigma_{ki}=1}^2\mu'_{k\to \Psi_i}(\sigma_{ki},\tau_i^{(k)},c_{ki}=1,t_i^{(k)})\right]\\
	-&a(t_i^{(j)}-t_j^{(i)}-1)\delta_{x_i^0,S}\mathbb{I}[\tau_i^{(j)}<{T+1}]\mathbb{I}[\sigma_{ji}=2]\prod_{k\in\partial i\setminus j}\mu'_{k\to \Psi_i}(\sigma_{ki}=2,\tau_i^{(k)},c_{ki}=1,t_i^{(k)})\\
	-&\phi(t_i^{(j)})a(t_i^{(j)}-t_j^{(i)})\delta_{x_i^0,I}\delta_{\tau_i^{(j)},0}\prod_{k\in\partial i\setminus j}\left[\sum_{\sigma_{ki}=0}^2\mu'_{k\to \Psi_i}(\sigma_{ki},\tau_i^{(k)},c_{ki}=0,t_i^{(k)})\right]\\
	-&\phi(t_i^{(j)})a(t_i^{(j)}-t_j^{(i)})\delta_{x_i^0,S}\mathbb{I}[\sigma_{ji}\in\{1,2\}]\prod_{k\in\partial i\setminus j}\left[\sum_{\sigma_{ki}=1}^2\mu'_{k\to \Psi_i}(\sigma_{ki},\tau_i^{(k)},c_{ki}=0,t_i^{(k)})\right]\\
	+&\left.\phi(t_i^{(j)})a(t_i^{(j)}-t_j^{(i)})\delta_{x_i^0,S}\mathbb{I}[\tau_i^{(j)}<{T+1}]\mathbb{I}[\sigma_{ji}=2]\prod_{k\in\partial i\setminus j}\mu'_{k\to \Psi_i}(\sigma_{ki}=2,\tau_i^{(k)},c_{ki}=0,t_i^{(k)})\right)
\end{aligned}
\end{align}
The second BP equation becomes:
\begin{align}
\label{eq:BP_variable_to_factor}
\left\{
\begin{array}{llllll}
	\mu'(0,\tau_j^{(i)},c_{ij},t_j^{(i)})&=\mathbb{I}[\tau_j-s_{ji}>0]\sum_{t_i^{(j)}}a(t_j^{(i)}-t_i^{(i)}-c_{ij})\sum_{\tau_i^{(j)}=0}^{\tau_j^{(i)}-s_{ji}}\tilde{\nu}_{\Psi_i\to j}(\tau_i^{(j)},\sigma_{ji}=2,t_i^{(j)},t_j^{(i)}) \\
	\mu'(1,\tau_j^{(i)},c_{ij},t_j^{(i)})&= \mathbb{I}[\tau_j-s_{ji}\geq 0]\sum_{t_i^{(j)}}a(t_j^{(i)}-t_i^{(i)}-c_{ij})\tilde{\nu}_{\Psi_i\to j}(\tau_i^{(j)}=\tau_j^{(i)}-s_{ji},\sigma_{ji}=2,t_i^{(j)},t_j^{(i)})\\	
	\mu'(2,\tau_j^{(i)},c_{ij},t_j^{(i)})&=\sum_{t_i^{(j)}}a(t_j^{(i)}-t_i^{(i)}-c_{ij})\left[\sum_{\tau_i^{(j)}=\zeta^+_i}^{\zeta^-_i}\tilde{\nu}_{\Psi_i\to j}(\tau_i^{(j)},\sigma_{ji}=2,t_i^{(j)},t_j^{(i)})\right. \\
	&+ \mathbb{I}[\tau_j^{(i)}+s_{ji}\leq T+1]\tilde{\nu}_{\Psi_i\to j}(\tau_i^{(j)}=\tau_j^{(i)}+s_{ji},\sigma_{ji}=1,t_i^{(j)},t_j^{(i)})\\
	&+\left. \mathbb{I}[\tau_j^{(i)}+s_{ji}<T+1]\sum_{\tau_i^{(j)}=\tau_j^{(i)}+s_{ji}+1}^{T+1} \tilde{\nu}_{\Psi_i\to j}(\tau_i^{(j)},\sigma_{ji}=0,t_i^{(j)},t_j^{(i)})\right]	
\end{array}
\right.
\end{align}

\subsection{BP marginals}
Once a fixed-point of the BP equations~(\ref{eq:BP_factor_to_variable},\ref{eq:BP_variable_to_factor}) is found, the BP marginal can be expressed as:
\begin{align}
\label{eq:BP_marginal}
\begin{aligned}
	P_i(\tau_i, t_i) &= \sum_{\underline{\tau}_{\partial i},\underline{t}_{\partial i}}P_{\Psi_i}(\tau_i, t_i, \underline{\tau}_{\partial i},\underline{t}_{\partial i}) \\
	&= \frac{1}{Z_{\Psi_i}}\sum_{\underline{\tau}_{\partial i},\underline{t}_{\partial i}}\xi(\tau_i, t_i;c_i)\psi^*(\tau_i, \underline{\tau}_{\partial_i};\{s_{li}\}_{l\in\partial i},x_i^0)\psi(t_i,\underline{t}_{\partial i})\prod_{l\in\partial i}\mu_{l\to\Psi_i}(\tau_l,\tau_i, t_l, t_i) \\
	&= \frac{1}{Z_{\Psi_i}}\gamma(t_i)\xi(\tau_i, t_i; c_i) \left(\delta_{x_i^0,I}\delta_{\tau_i,0}\prod_{l\in\partial i}\left[\sum_{\sigma_{li}=0}^2\mu'_{l\to\Psi_i}(\sigma_{li}, \tau_i, c_{li}=1, t_i)\right]\right. \\
	+& \delta_{x_i^0,S}\prod_{l\in\partial i}\left[\sum_{\sigma_{li}=1}^2\mu'_{l\to \Psi_i}(\sigma_{li}, \tau_i, c_{li}=1, t_i)\right] \\
	-& \delta_{x_i^0,S}\mathbb{I}[\tau_i<{T+1}]\prod_{l\in\partial i}\mu_{l\to \Psi_i}(\sigma_{li}=2, \tau_i, c_{li}=1, t_i) \\
	-&\phi(t_i)\delta_{x_i^0,I}\delta_{\tau_i,0}\prod_{l\in\partial i}\left[\sum_{\sigma_{li}=0}^2\mu'_{l\to\Psi_i}(\sigma_{li}, \tau_i, c_{li}=0, t_i)\right]\\
	-&\phi(t_i)\delta_{x_i^0,S}\prod_{l\in\partial i}\left[\sum_{\sigma_{li}=1}^2\mu'_{l\to \Psi_i}(\sigma_{li}, \tau_i, c_{li}=0, t_i)\right]\\
	+&\left.\phi(t_i) \delta_{x_i^0,S}\mathbb{I}[\tau_i<{T+1}]\prod_{l\in\partial i}\mu'_{l\to \Psi_i}(\sigma_{li}=2, \tau_i, c_{li}=0, t_i)\right)
\end{aligned}
\end{align}

\subsection{Bethe Free Energy}
\label{subsec:BetheFreeEnergy}
The Bethe Free energy is written:
\begin{align}
\begin{aligned}
F = -\sum_{i\in V}\log Z_{\Psi_i} + \frac{1}{2}\sum_{i\in V}\sum_{j\in\partial i}\log Z_{ij}
\end{aligned}
\end{align}
where $Z_{\Psi_i}$ is the normalisation of the BP marginal written above, and with:
\begin{align}
\begin{aligned}
Z_{ij} &= \sum_{T_{ij}}\nu_{\Psi_i\to j}(T_{ij})\nu_{\Psi_j\to i}(T_{ij}) \\
&= \sum_{T_{ij}}\nu_{\Psi_i\to j}(T_{ij})\mu_{j\to \Psi_i}(T_{ij}) \\
&= \frac{1}{z_{\Psi_i\to j}}\sum_{\{T_{il}\}_{l\in\partial i}}\Psi(\{T_{il}\}_{l\in\partial i})\prod_{l\in\partial i}\mu_{l\to \Psi_i}(T_{il}) \\
&= \frac{Z_{\Psi_i}}{z_{\Psi_i\to j}}
\end{aligned}
\end{align}
Where $z_{\Psi_i\to j}$ is the normalization of the BP message $\nu_{\Psi_i\to j}$:
\begin{align}
\begin{aligned}
z_{\Psi_i\to j} &= \sum_{\{T_{il}\}_{l\in\partial i}}\Psi(\{T_{il}\}_{l\in\partial i})\prod_{k\in\partial i\setminus j}\mu_{k\to \Psi_i}(T_{ik}) \\
&= \sum_{\tau_i^{(j)},\tau_j^{(i)},t_i^{(j)},t_j^{(i)}}\tilde{\nu}(\tau_i^{(j)},\sigma_{ji}=1+{\rm sgn}(\tau_j^{(i)}+s_{ji}-\tau_i^{(j)}), t_i^{(j)},t_j^{(i)})
\end{aligned}
\end{align}
where $\tilde{\nu}$ is the un-normalized message defined in~(\ref{eq:tilde_nu}).
We obtain an expression of the free-energy in terms of the normalisations $Z_{\Psi_i}$, $z_{\Psi_i\to j}$:
\begin{align}
F = \frac{1}{N}\sum_{i\in V}\left(\frac{d_i}{2}-1\right)\log Z_{\Psi_i} - \frac{1}{N}\frac{1}{2}\sum_{i\in V}\sum_{j\in\partial i}\log z_{\Psi_i\to j}
\end{align}
\subsection{Entropy and Energy}

To compute the entropy it is sufficient to subtract energy and free
energy:
\[
S=U-F
\]
The energy is simply the average of the Hamiltonian:
\begin{align*}
H & =-\sum_{i}\log\psi-\sum_{i}\log\xi-\sum_{i}\log\psi^{*}\\
U & =-\sum_{i}\left\langle \log\psi\right\rangle -\sum_{i}\left\langle \log\xi\right\rangle -\sum_{i}\left\langle \log\psi^{*}\right\rangle =\\
 & =-\sum_{i}\left\langle \log\psi\right\rangle .
\end{align*}
\[
U=-\sum_{i}\frac{1}{Z_{\Psi_{i}}}u_{\Psi_{i}},
\]
with:
\[
u_{\Psi_{i}}:=\sum_{\{T_{ij}\}_{j\in\partial i}}\prod_{j\in\partial i}m_{j\to\Psi_{i}}(T_{ij})\psi(t_{i},t_{\partial i})\psi^{*}(\tau_{i},\tau_{\partial i})\xi(t_{i},\tau_{i})\log\psi(t_{i},t_{\partial i}).
\]
Comparing this formula with the expression for $Z_{\Psi_{i}}$:
\[
Z_{\Psi_{i}}=\sum_{\{T_{ij}\}_{j\in\partial i}}\prod_{j\in\partial i}m_{j\to\Psi_{i}}(T_{ij})\psi^{*}(\tau_{i},\tau_{\partial i})\xi(t_{i},\tau_{i})\psi(t_{i},t_{\partial i})
\]
we see that the computation of the energy requires similar calculations
to the ones already performed to compute free energy. The only additional
difficulty is in the $\log\psi$ factor. Let us first trace over the
planted variables. We keep $\tau_{i}$ fixed:
\begin{align*}
 & \sum_{\tau_{\partial i}}\prod_{j\in\partial i}m_{j\to\Psi_{i}}(T_{ij})\psi^{*}(\tau_{i},\tau_{\partial i})=\\
= & \delta_{x_{i,0}^{*},I}\delta_{\tau_{i},0}\prod_{j\in\partial i}\sum_{\tau_{j}}m_{\Psi_{j}\to i}(t_{j},t_{i},\tau_{j},\tau_{i})\\
 & +\delta_{x_{i,0}^{*},S}\prod_{j\in\partial i}\sum_{\tau_{j}}m_{\Psi_{j}\to i}(t_{j},t_{i},\tau_{j},\tau_{i})\mathbb{I}[\tau_{i}\leq\tau_{j}+s_{ji}]\\
 & -\mathbb{I}[\tau_i\leq T]\delta_{x_{i,0}^{*},S}\prod_{j\in\partial i}\sum_{\tau_{j}}m_{\Psi_{j}\to i}(t_{j},t_{i},\tau_{j},\tau_{i})\mathbb{I}[\tau_{i}<\tau_{j}+s_{ji}]=\\
=: & \sum_{v=1}^{3}\prod_{j\in\partial i}m_{j\to\Psi_{i}}^{v}(t_{i},t_{j},\tau_{i})
\end{align*}
So we have:
\[
Z_{\Psi_{i}}=\sum_{t_{i},\tau_{i},\{t_{j}\}_{j\in\partial i}}\sum_{v=1}^{3}f_v^{\tau_i,x_{i,0}^*}\prod_{j\in\partial i}m_{j\to\Psi_{i}}^{v}(t_{i},t_{j},\tau_{i})\xi(t_{i},\tau_{i})\psi(t_{i},t_{\partial i})
\]
Now remember that the original message $m$ and the compressed factor
to node message $\nu$ are related by :
\[
\nu_{\psi_{i}\to j}(t_{i},t_{j},\tau_{i},1+\text{sign}(\tau_{j}-\tau_{i}+s_{ji}))=m_{\psi_{i}\to j}(t_{i},t_{j},\tau_{i},\tau_{j})
\]
So the sums we have to compute are:
\begin{align*}
m_{j\to\Psi_{i}}^{0}(t_{i},t_{j},0) & =\sum_{\tau_{j}}m_{\Psi_{j}\to i}(t_{j},t_{i},\tau_{j},0)\\
 & =\sum_{\tau_{j}}\nu_{\psi_{j}\to i}(t_{j},t_{i},\tau_{j},1+\text{sign}(-\tau_{j}+s_{ij}))\\
m_{j\to\Psi_{i}}^{1}(t_{i},t_{j},\tau_{i}) & =\sum_{\tau_{j}\geq\tau_{i}-s_{ji}}m_{\Psi_{j}\to i}(t_{j},t_{i},\tau_{j},\tau_{i})=\\
 & =\sum_{\tau_{j}\geq\tau_{i}-s_{ji}}\nu_{\Psi_{j}\to i}(t_{j},t_{i},\tau_{j},1+\text{sign}(\tau_{i}-\tau_{j}+s_{ij}))\\
m_{j\to\Psi_{i}}^{2}(t_{i},t_{j},\tau_{i}) & =\sum_{\tau_{j}>\tau_{i}-s_{ji}}m_{\Psi_{j}\to i}(t_{j},t_{i},\tau_{j},\tau_{i})=\\
 & =\sum_{\tau_{j}>\tau_{i}-s_{ji}}\nu_{\Psi_{j}\to i}(t_{j},t_{i},\tau_{j},1+\text{sign}(\tau_{i}-\tau_{j}+s_{ij}))
\end{align*}
Notice that:
\[
m_{j\to\Psi_{i}}^{1}(t_{i},t_{j},\tau_{i})=m_{j\to\Psi_{i}}^{2}(t_{i},t_{j},\tau_{i})+\nu_{\Psi_{j}\to i}(t_{j},t_{i},\tau_{i}-s_{ji},2).
\]
Now it is time to deal with planted times. Let us observe that:
\[
\psi(t_{i},t_{\partial i})=\gamma(t_{i})(1-\lambda)^{S_{1}}\left(1-(1\leq t_{i}\leq T)(1-\lambda)^{S_{2}}\right),
\]
where:
\begin{align*}
\gamma(t_{i}) & =\gamma\delta_{t_{i},0}+(1-\gamma)\left(1-\delta_{t_{i},0}\right)\\
S_{1}:= & \sum_{j\in\partial i}(t_{i}-t_{j}-1)_{+}\\
S_{2}:= & \sum_{j\in\partial i}\theta(t_{i}-t_{j}-1)
\end{align*}
We want to find the BP distribution of $t_{i},S_{1},S_{2}$ in order
to average over $\psi\log\psi.$ We define:
\begin{align*}
F_{k}^{v,\tau_{i}}\left(t_{i},S_{1},S_{2}\right): & =\sum_{\{t_{j}\}_{j\leq k}}\prod_{j\leq k}m_{j\to\Psi_{i}}^{v}(t_{i},t_{j},\tau_{i})\delta_{S_{1},\sum_{j\leq k}(t_{i}-t_{j}-1)_{+}}\delta_{S_{2},\sum_{j\leq k}\theta(t_{i}-t_{j}-1)}
\end{align*}
Therefore:
\begin{align*}
F_{k+1}^{v,\tau_{i}}\left(t_{i},S_{1},S_{2}\right) & =\sum_{\{t_{j}\}_{j\leq k+1}}\prod_{j\leq k+1}m_{j\to\Psi_{i}}^{v}(t_{i},t_{j},\tau_{i})\delta_{S_{1},\sum_{j\leq k+1}(t_{i}-t_{j}-1)_{+}}\delta_{S_{2},\sum_{j\leq k+1}\theta(t_{i}-t_{j}-1)}=\\
 & =\sum_{t_{k+1}}m_{k+1\to\Psi_{i}}^{v}(t_{i},t_{k+1},\tau_{i})\sum_{\{t_{j}\}_{j\leq k}}\prod_{j\leq k}m_{j\to\Psi_{i}}^{v}(t_{i},t_{j},\tau_{i})\times\\
 & \qquad\qquad\times\delta_{S_{1}-(t_{i}-t_{k+1}-1)_{+},\sum_{j\leq k}(t_{i}-t_{j}-1)_{+}}\delta_{S_{2}-\theta(t_{i}-t_{k+1}-1),\sum_{j\leq k}\theta(t_{i}-t_{j}-1)}=\\
 & =\sum_{t_{k+1}}m_{k+1\to\Psi_{i}}^{v}(t_{i},t_{k+1},\tau_{i})F_{k}^{v,\tau_{i}}(t_{i},S_{1}-(t_{i}-t_{k+1}-1)_{+},S_{2}-\theta(t_{i}-t_{k+1}-1))
\end{align*}
and 
\[
F_{0}^{v,\tau_{i}}(t_{i},S_{1},S_{2})=\delta_{S_{1},0}\delta_{S_{2},0}
\]
and for our purposes we want to find $F_{|\partial i|}^{v,\tau_{i}}(t_{i},S_{1},S_{2})$.
Now we have an iterative scheme to compute the measure. Once the function
is found we simply have:
\begin{align*}
u_{\Psi_{i}} & =\sum_{t_{i},\tau_{i}:\xi(t_{i},\tau_{i})=1}\sum_{v=1}^{3}f_v^{\tau_i,x_{i,0}^*}\sum_{S_{1},S_{2}}F_{|\partial i|}^{v,\tau_{i}}(t_{i},S_{1},S_{2})\psi(t_{i},S_{1},S_{2})\log\psi(t_{i},S_{1},S_{2})
\end{align*}

\section{Replica Symmetric Formalism} 
\label{sec:RS_equations}
The aim of the cavity method is to characterize the typical properties of the probability measure~(\ref{eq:prob_auxiliary}) that we recall here:
\begin{align*}
P(\{T_{ij}\}_{(ij)\in E}|\mathcal{D}) &= \frac{1}{\mathcal{Z}(\mathcal{D})}\prod_{i\in V}\Psi(\{T_{il}\}_{l\in\partial i};\mathcal{D}_i) \ ,
\end{align*}
for typical random graphs and for typical realization of the disorder $\mathcal{D}=\{\mathcal{D}_i\}$, in the thermodynamic limit $N\to\infty$.
In the simplest version of the cavity method, called Replica Symmetric (RS), one assumes a fast decay of the correlations between distant variables in the measure (\ref{eq:prob_auxiliary}), in such a way that the BP equations~(\ref{eq:BP_equations}) converge to a unique fixed-point on a typical large instance, and that the measure (\ref{eq:prob_auxiliary}) is well described by the locally tree-like approximation.
We consider a uniformly chosen edge $(ij)\in E$ in a random contact graph $\mathcal{G}$, and call $\mathcal{P}^{\rm rs}$ the probability law of the fixed-point BP message $\mu_{i\to \Psi_j}$ thus observed.
Within the decorrelation hypothesis of the RS cavity method, the incoming messages on a given factor node are $i.i.d.$ with probability $\mathcal{P}^{\rm rs}$.
This implies that the probability law $\mathcal{P}^{\rm rs}$ must obey the following self-consistent equation:
	\begin{align}
	\mathcal{P}^{\rm rs}(\mu) = \sum_{d=0}^\infty r_d\sum_{\mathcal{D}_i}P(\mathcal{D}_i)\int\prod_{i=1}^d{\rm d}p^{\rm rs}(\mu_i)\delta(\mu - f^{\rm bp}(\mu_1,\dots,\mu_d;\mathcal{D}_i))
	\end{align}
where $f^{\rm bp}(\mu_1,\dots,\mu_d;\mathcal{D}_i)$ is a shorthand notation for the r.h.s. of equation (\ref{eq:BP_equations}), and $p(\mathcal{D}_i)$ is distribution of the local disorder $\mathcal{D}_i=\{\{s_{li}\}_{l\in\partial i},x_i^0, \{\varepsilon_m\}_{i_m=i} \}$ associated with a given node $i$.
We numerically solved these equations with population dynamics.
Using the above simplifications, we are left with two types of BP messages: $\mu'_{i\to\Psi_j}$ is defined over the variable $((\sigma_{ij},\tau_j^{(i)},c_{ij},t_j^{(i)}))$ living in a space of size $6(T+1)^2$, and $\tilde{\nu}_{\Psi_i\to j}$ is defined over the variable $(\tau_i^{(j)},\sigma_{ji},t_i^{(j)},t_j^{(i)})$, living in a space of size $3(T+1)^3$.
We store only a population of messages $\mu_{i\to \Psi_j}$, this requires to keep in memory $O(\mathcal{N}T^2)$ numbers, with $\mathcal{N}$ the population size.
Computing a new element $\mu$ of the population requires in principle $O(T^4)$ operations, but can be reduced to $O(T^3)$ by computing the cumulants of the temporary message $\nu$.

\subsection{Replica-Symmetric Free Energy}
Once averaged over the graph and disorder, the Replica Symmetric prediction for the free-energy is:
\begin{align}
\begin{aligned}
F^{\rm RS} &= \sum_d p_d \left(\frac{d}{2}-1\right)\sum_cp(c)\sum_{x}\gamma(x)\prod_{i=1}^dw(s_i)\int\prod_{l=1}^d{\rm d}\mathcal{P}^{\rm RS}(\mu_l)\log Z_{\Psi_i}(\mu_1,\dots,\mu_d;x,c,s_1,\dots,s_d) \\
&-\frac{d_{\rm av}}{2}\sum_dr_d\sum_{x}\gamma(x)\sum_cp_c\sum_{s_1,\dots,s_d}\prod_{k=1}^dw(s_k)\int\prod_{k=1}^d{\rm d}\mathcal{P}^{\rm RS}(\mu_k)\log z_{\Psi_i\to j}(\mu_1,\dots,\mu_d;x,c,s_1,\dots,s_d)
\end{aligned}
\end{align}
\end{document}